\documentclass[lettersize,journal]{IEEEtran}

\usepackage{color}
\usepackage{amsmath,amsthm,amssymb,amsfonts}

\allowdisplaybreaks[4]

\usepackage[T1]{fontenc}
\usepackage{graphicx}
\usepackage{longtable}
\usepackage{float}
\usepackage{hyperref}

\usepackage{bm}

\usepackage[bottom]{footmisc}

\usepackage[mathscr]{eucal}

\usepackage[caption=false,font=normalsize,labelfont=sf,textfont=sf]{subfig}

\usepackage{multirow}
\usepackage{makecell}

\usepackage{cite}

\usepackage{epsfig}

\usepackage{soul}

\usepackage[lined,boxed,ruled]{algorithm2e}

\newcommand{\barjmath}{\bar{\jmath}}
\newtheorem{theorem}{Theorem}
\newtheorem{lemma}{Lemma}

\newcommand{\Exp}{{\mathbb{E}}}

\newcommand{\Var}[1]{\mathbb{V}\braces{#1}}

\newcommand{\braces}[1]{\left\lbrace #1\right\rbrace}

\newcommand{\Dkl}[2]{D_{\textrm{KL}}\left\lbrace #1 : #2\right\rbrace}

\newcommand{\setposi}[1]{\mathcal{Z}_{#1}^+}

\newcommand{\setnnega}[1]{\mathcal{Z}_{#1}}

\newcommand{\Diag}[1]{\mathrm{Diag}\left\lbrace #1\right\rbrace }

\newcommand{\argmax}[1]{\mathop{\arg\max}\limits_{#1}}
\newcommand{\argmin}[1]{\mathop{\arg\min}\limits_{#1}}

\newcommand{\ba}{\mathbf{a}}

\newcommand{\bd}{\mathbf{d}}

\newcommand{\bg}{\mathbf{g}}
\newcommand{\bh}{\mathbf{h}}

\newcommand{\bs}{\mathbf{s}}
\newcommand{\bt}{\mathbf{t}}

\newcommand{\bv}{\mathbf{v}}
\newcommand{\bw}{\mathbf{w}}
\newcommand{\bx}{\mathbf{x}}
\newcommand{\by}{\mathbf{y}}
\newcommand{\bz}{\mathbf{z}}

\newcommand{\bA}{\mathbf{A}}

\newcommand{\bG}{\mathbf{G}}

\newcommand{\bI}{\mathbf{I}}

\newcommand{\bU}{\mathbf{U}}

\newcommand{\bbC}{\mathbb{C}}
\newcommand{\bbR}{\mathbb{R}}

\newcommand{\bzero}{\mathbf{0}}

\newcommand{\bSigma}{{\boldsymbol\Sigma}}

\newcommand{\bLambda}{{\boldsymbol\Lambda}}

\newcommand{\btheta}{{\boldsymbol\theta}}

\newcommand{\bmu}{{\boldsymbol\mu}}

\newcommand{\bxi}{{\boldsymbol\xi}}

\newcommand{\normmm}[1]{{\left\vert\kern-0.25ex\left\vert\kern-0.25ex\left\vert #1 
		\right\vert\kern-0.25ex\right\vert\kern-0.25ex\right\vert}}

\newcommand{\norm}[1]{\lVert #1 \rVert}

\definecolor{myback}{RGB}{204,232,207}

\usepackage{makecell}

\begin{document}

\title{Group Information Geometry Approach for Ultra-Massive MIMO Signal Detection}	
\author{
	Jiyuan~Yang,~\IEEEmembership{~Member,~IEEE,}
	~Yan~Chen, ~Mingrui~Fan,
    ~Xiqi~Gao,~\IEEEmembership{Fellow,~IEEE,}
    Xiang-Gen~Xia,~\IEEEmembership{Fellow,~IEEE,}
    and~Dirk~Slock,~\IEEEmembership{Fellow,~IEEE} 
}

\maketitle
\begin{abstract}
We propose a group information geometry approach (GIGA) for ultra-massive multiple-input multiple-output (MIMO) signal detection. 
The signal detection task is framed as computing the approximate marginals of the {\textsl{a posteriori}} distribution of the transmitted data symbols of all users.
With the approximate marginals, we perform the maximization of the {\textsl{a posteriori}} marginals (MPM) detection to recover the symbol of each user.
Based on the information geometry theory and the grouping of the components of the received signal, three types of manifolds are constructed and the approximate  {\textsl{a posteriori}} marginals are obtained through  \textsl{m}-projections.
The Berry-Esseen theorem is introduced to offer an approximate calculation of the \textsl{m}-projection, while its direct calculation is exponentially complex.
In most cases, more groups, less complexity of GIGA.
However, when the number of groups exceeds a certain threshold, the complexity of GIGA starts to increase.
Simulation results confirm that the proposed GIGA achieves better bit error rate (BER) performance within a small number of iterations, which demonstrates that it can serve as an efficient detection method in ultra-massive MIMO systems.
\end{abstract}

\section{Introduction}

Emerging as a promising solution to address the capacity demands of future communication systems, ultra-massive multiple-input multiple-output (MIMO)  (also known as extremely large-scale MIMO,  extra-large scale massive MIMO, etc.) has attracted a significant attention.
Ultra-massive MIMO leverages a substantial number of antennas at the base station (BS), often hundreds to thousands, to serve a large number of user terminals on the same time-frequency resource, which can dramatically improve spectrum efficiency, energy efficiency, and spatial resolution \cite{9693928,9547795,9617121,9551696,9957130,10078317}.
This technology also offers substantial beamforming gains, crucial for mitigating path losses at high-frequency bands like millimeter-wave (mmWave) and terahertz (THz) \cite{9957130,8901159}. 

To fulfill the various advantages of ultra-massive MIMO, signal detection plays an important role.
Typically, signal detection is used to recover the transmitted symbols of the user terminals based on the received signal and the channel state information (CSI).
In MIMO transmission, inter-symbol interference and noise pose a great challenge for signal detection.
In general, the maximum \textsl{a posteriori} (MAP) or maximum-likelihood (ML) detection could provide a statistically optimal solution by means of an exhaustive search over all possible transmitted symbols.
Nevertheless, the combinatorial nature of the MAP or ML detection makes conventional numerical algorithms for convex optimization unsuitable.
The MAP or ML detection can be prohibitively complex in practice.
On the other hand, classical linear detectors such as  zero-forcing (ZF) and linear minimum mean-squared error (LMMSE) detectors, suffer from limited performance despite their polynomial-time complexity \cite{6841617,7244171,9159940}.

The massive MIMO signal detection has been a topic of great interest during the past few years.
Numerous algorithms have been proposed to address this problem \cite{9250659,6415398,9484686,9139393,5961820}.
Readers interested in the development of massive MIMO signal detection techniques can refer to \cite{8804165}.
Among all the signal detection algorithms, those based on Bayesian inference, such as belief propagation
(BP), expectation propagation (EP), and approximate message propagation (AMP), have gained a lot of interest due to their satisfactory performance-complexity profile.
While possessing reduced complexity compared to MAP or ML detection, these algorithms can still achieve sub-optimal detection performance.
Based on the Markov random field and message passing, a low-complexity detection algorithm is proposed for large dimensional MIMO-ISI channels in \cite{6008619}.
The paper \cite{6841617} pioneers the integration of EP into massive MIMO signal detection with high-order modulation schemes.
In \cite{6778065}, an iterative detector based on AMP is proposed for large-scale multiuser MIMO-OFDM systems.

Recently, an interdisciplinary field, information geometry, has attracted great interest.
It merges concepts from information theory and differential geometry to explore the geometric structures and properties of statistical models \cite{IGanditsAPP,srbpig,amari}.
From the perspective of information geometry, the set of distributions can be represented as a manifold, offering a natural approach to describe the relationship between different sets of probability distributions.
One common metric used to measure the distance between different probability distributions is the well-known Kullback-Leibler (K-L) divergence.
Information geometry provides a mathematical foundation for analyzing and understanding the intrinsic geometric structures of various statistical models.
As an active and powerful subject, it has been widely used in research related to statistics, such as radar target detection \cite{10399943,9479799}, channel estimation \cite{IGA}, and quantitative fault diagnosability analysis \cite{9741345}.

Recently, we have proposed an information geometry approach (IGA) for ultra-massive MIMO signal detection in \cite{IGADE}.
In \cite{IGADE}, the signal detection problem is framed as computing the approximate marginals of the \textsl{a posteriori} probability distribution of the transmitted symbols.
Based on the \textsl{a posteriori} probability distribution, we define the objective manifold (OBM) and the auxiliary manifolds (AMs), where the OBM contains the approximate marginal probability distributions and each AM is related to the received signal on a single antenna at the receiver.
The calculation of the approximation is then converted to  $m$-projections from the distributions of AMs onto the OBM.
Although it can provide high detection performance, IGA has a slow convergence rate when modulation order and signal-to-noise ratio are high.

In this work, we propose a group information geometry approach (GIGA) for signal detection of ultra-massive MIMO systems.
Our goal is still to acquire the approximation of the \textsl{a posteriori} marginals, and the maximization of the a posteriori
marginals (MPM) detection is performed to recover the transmitted data symbols from the perspective of information geometry.
Different from the IGA in \cite{IGADE}, we express the \textsl{a posteriori} distribution in a factorized manner by grouping the components of the received signal.
On this basis, we define new AMs, where each AM is related to the received signal on a group of antennas at the receiver.
The approximate marginals are then obtained through  $m$-projections from the distributions of AMs onto the OBM.
A direct calculation of the $m$-projection is first presented, whose complexity is exponential and unaffordable.
To solve this problem, we propose an approximate calculation of the $m$-projection based on the Berry-Esseen theorem, which can reduce the complexity significantly. 
Simulation results show that GIGA has significant advantages over existing algorithms in both convergence rate and detection performance.
In general, the complexity of GIGA decreases as the number of groups increases. 
When the number of groups exceeds a certain threshold, its complexity starts to increase.
Given a proper number of groups, GIGA can obtain better BER performance with lower computational complexity compared to IGA.

The remaining sections are arranged as follows. 
Section \uppercase\expandafter{\romannumeral2} introduces the system configuration and problem formulation.
GIGA is developed in Section \uppercase\expandafter{\romannumeral3}. 
The calculation of the $m$-projection is discussed in Section \uppercase\expandafter{\romannumeral4}. 
Simulation results are discussed in Section \uppercase\expandafter{\romannumeral5}.
Finally, conclusions are drawn in Section \uppercase\expandafter{\romannumeral6}.

Throughout this paper, upper (lower) case boldface letters denote matrices
(column vectors). 
$\Exp_p\left\lbrace \cdot \right\rbrace $ denotes the expectation operation w.r.t. the distribution $p\left(\bh\right)$.  
$\mathcal{R}\left(\cdot\right)$ and $\mathcal{I}\left(\cdot\right)$ denote the real and imaginary parts of a complex matrix, respectively.
We use $y_n$ or $\left[\by\right]_n$,
$a_{i,j}$ or $\left[\bA\right]_{i,j}$ to denote the $n$-th component of the vector $\by$ and the $\left(i,j\right)$-th component of the matrix $\bA$, where the element indices start with $1$. 
Given a vector $\by \in \bbC^{P}$, where $P \ge 1$ is an integer. 
Define a set $\mathcal{B} \triangleq \braces{b_1, b_2, \cdots, b_Q}$, where $1\le b_1 < b_2 < \cdots < b_Q \le P$ and $Q \le P$.
We denote $\left[\by\right]_{\mathcal{B}}$ as 
\begin{equation*}
	\left[\by\right]_{\mathcal{B}} \triangleq \left[ \left[\by\right]_{b_1}, \left[\by\right]_{b_2}, \cdots, \left[ \by \right]_{b_Q}  \right]^T \in \bbC^{Q}.
\end{equation*}
Given a matrix $\bA \in \bbC^{P \times P'}$, we use
$\left[ \bA \right]_{m,:}$ to denote the $m$-th row of the matrix $\bA$.
We denote $\left[ \bA \right]_{\mathcal{B},:}$ as
\begin{equation*}
	\left[ \bA \right]_{\mathcal{B},:} \triangleq \left[ \left[ \bA \right]_{b_1,:}^T, \left[ \bA \right]_{b_2,:}^T, \cdots, \left[ \bA \right]_{b_Q,:}^T   \right]^T \in \bbC^{Q\times P'}.
\end{equation*}
$\odot$ and $\otimes$ denote the Hadamard product and  Kronecker product, respectively. 
We define $\setnnega{N} \triangleq \braces{0,1,\ldots,N}$ and $\setposi{N} \triangleq \braces{1,2,\ldots,N}$.
$\norm{\cdot}$ denotes the $\ell_2$ norm.
We use $p\left(\cdot\right)$ to represent the probability distribution of discrete random variables and $f\left(\cdot\right)$ to represent the probability density function (PDF) of continuous random variables, respectively.
We denote the PDF of a complex Gaussian random vector $\bx\sim\mathcal{CN}\left( \bmu,\bSigma \right)$ as
$f_{\textrm{CG}}\left( \bx;\bmu,\bSigma \right)$.
We denote the PDF of a real Gaussian vector $\bx \sim \mathcal{N}\left( \bmu,\bSigma \right)$ as $f_{\textrm{G}}\left( \bx;\bmu,\bSigma \right)$.
Given a positive-definite matrix $\bA \in \bbC^{m\times m}$, then $\bA$
can be decomposed as $\bA = \bU\bLambda\bU^H$, where $\bU \in \bbC^{m\times m}$ is unitary,  $\bLambda \triangleq \Diag{\lambda_1, \lambda_2, \ldots, \lambda_m}$, and  $\braces{\lambda_i}_{i=1}^m$ are all the eigenvalues of $\bA$.
We denote $\bA^{-\frac{1}{2}}$ as $\bA^{-\frac{1}{2}} \triangleq \bU^H\bLambda^{-\frac{1}{2}}\bU$, where $\bLambda^{-\frac{1}{2}} \triangleq \Diag{\sqrt{\lambda_1}, \sqrt{\lambda_2}, \ldots, \sqrt{\lambda_m}}$.
Define the delta function $\delta\left(x - c\right)$ as 
\begin{equation}
  \delta\left(x-c\right)=
	\begin{cases}
		1, &\textrm{when}\quad  x = c,\\
		0, &\textrm{otherwise},
	\end{cases}
\end{equation}
where $c$ is a constant.


\section{System Configuration and Problem Formulation}
In this section, we give the configuration of the considered ultra-massive MIMO system.
Then, we present the problem formulation of the ultra-massive MIMO signal detection.

\subsection{System Configuration}
Consider an uplink ultra-massive MIMO transmission where $K$ single-antenna users are served by a base station (BS) with an ultra-massive antenna array of $N_r$ antennas.
Denote the transmitted data symbol of user $k$ as $\tilde{s}_k \in \tilde{\mathbb{S}}$, where 
\begin{equation*}
	\tilde{\mathbb{S}} \triangleq \braces{\tilde{s}^{\left(0\right)},\tilde{s}^{\left(1\right)},\ldots,\tilde{s}^{\left(\tilde{L}-1\right)}}
\end{equation*}
is the signal constellation, $\braces{\tilde{s}^{\left(\ell\right)}}_{\ell=0}^{\tilde{L}-1}$ are the constellation points, and $\tilde{L}$ is the  modulation order (or constellation size).
Throughout this work, our focus is on uncoded systems employing symmetric $\tilde{L}$-QAM modulation. 
We assume that each user selects symbols uniformly from $\tilde{\mathbb{S}}$, and all users share the same signal constellation\footnote{The proposed GIGA can be easily extended to any modulation with varying selecting probability, provided that the symbols of different users are statistically independent, and the real and imaginary parts of each user's symbol are statistically independent as well.}. 
In this paper, we also assume that the average power of each transmitted symbol is normalized to unit, i.e., $\Exp \braces{ \left|\tilde{s}_k\right|^2}  = 1$, $k\in \setposi{K}$.
Denote the transmitted symbol of all users as $\tilde{\bs} \triangleq \left[\tilde{s}_1,\tilde{s}_2,\ldots,\tilde{s}_K  \right]^T \in\tilde{\mathbb{S}}^K$.
We assume that  $\tilde{\bs}$ is transmitted through a flat-fading channel.
Then, the received signal $\tilde{\by}\in\bbC^{N_r}$ at the BS is given by 
\begin{equation}\label{equ:rece model 1}
	\tilde{\by} = \tilde{\bG}\tilde{\bs} + \tilde{\bz},
\end{equation}
where $\tilde{\bG} \in\bbC^{N_r\times K}$ is the channel matrix,  
$\tilde{\bz}$ is the circular-symmetric complex Gaussian noise, 
$\tilde{\bz} \sim\mathcal{CN}\left(\mathbf{0},\tilde{\sigma}_z^2\bI\right)$ 
and $\tilde{\sigma}_z^2$ is the noise variance\footnote{In the above notations, tildes are placed atop the mathematical symbols. 
This is to simplify the notation when formulating and analyzing their real counterparts later on, where the tildes are removed.}.
In this work, we assume perfect CSI at the BS.

\subsection{Problem Formulation}\label{sec:Probelm Statement}
Assuming that the noise vector $\tilde{\bz}$ and the transmitted symbol vector $\tilde{\bs}$ are independent, as are the symbols transmitted by different users. 
Given the received signal model \eqref{equ:rece model 1}, the \textsl{a posteriori} probability distribution of the transmitted symbol vector $\tilde{\bs}$ can be expressed as \cite{IGADE}
\begin{equation}
	\begin{split}
	   	p\left(\tilde{\bs}|\tilde{\by}\right) \propto\prod_{k=1}^{K}p_{\textrm{pr},k}\left(\tilde{s}_k\right)f_{\textrm{CG}}\left(\tilde{\by};\tilde{\bG}\tilde{\bs},\tilde{\sigma}_z^2\bI\right),
	\end{split}
\end{equation}
where 
$p_{\textrm{pr},k}\left(\tilde{s}_k\right)$ is the a priori probability of $\tilde{s}_k$, and
\begin{equation*}
	p_{\textrm{pr},k}\left(\tilde{s}_k\right)\big|_{\tilde{s}_k = \tilde{s}^{\left(\ell\right)}} = \frac{1}{\tilde{L}}, k\in \setposi{K}, \ell \in \setnnega{\tilde{L}-1}.
\end{equation*}
Given $p\left(\tilde{\bs}|\tilde{\by}\right)$, the  MAP detector (or, equivalently, the ML detector under this case) is  given by
\begin{equation}\label{equ:MAP}
	\tilde{\bs}_{\textrm{MAP}} = \argmax{\tilde{\bs}\in\tilde{\mathbb{S}}^K}p\left(\tilde{\bs}|\tilde{\by}\right).
\end{equation}
The optimization problem above is NP-hard due to the finite-alphabet constraint $\tilde{\bs}\in\tilde{\mathbb{S}}^K$.
When the number $K$ of users and the modulation order $\tilde{L}$ are large, the computation of \eqref{equ:MAP} will become unaffordable for practical applications.

In this paper, we process the real-valued counterpart of the 
received signal model in \eqref{equ:rece model 1}, which is essential for the development of GIGA.
To do so, let us first define the real-valued counterpart of \eqref{equ:rece model 1}.
Define real counterparts of $\tilde{\by}$, $\tilde{\bz}$, $\tilde{\bs}$ and  $\tilde{\bG}$ as 
\begin{subequations}
\begin{equation}\label{equ:real y and z}
	\by \triangleq \left[
	\begin{array}{*{20}{c}}
		\mathcal{R}\braces{\tilde{\by}} \\
		\mathcal{I}\braces{\tilde{\by}} 
	\end{array} 
	\right],
	\bz \triangleq \left[
	\begin{array}{*{20}{c}}
		\mathcal{R}\braces{\tilde{\bz}} \\
		\mathcal{I}\braces{\tilde{\bz}} 
	\end{array} 
	\right] 
	\in \bbR^{2N_r}
\end{equation}
\begin{equation}\label{equ:real s}
		\bs \triangleq \left[
	\begin{array}{*{20}{c}}
		\mathcal{R}\braces{\tilde{\bs}} \\
		\mathcal{I}\braces{\tilde{\bs}} 
	\end{array} 
	\right] 
	\in \bbR^{2K},
\end{equation}
\begin{equation}\label{equ:definition of G}
	\bG \triangleq \left[
	\begin{array}{*{20}{c}}
		\mathcal{R}\braces{\tilde{\bG}}, &-\mathcal{I}\braces{\tilde{\bG}}\\
		\mathcal{I}\braces{\tilde{\bG}} ,&\mathcal{R}\braces{\tilde{\bG}}
	\end{array} 
	\right]\in \bbR^{2N_r\times 2K},
\end{equation}
\end{subequations}
respectively.
Then, the real-valued counterpart of \eqref{equ:rece model 1} is given by
\begin{equation}\label{equ:rece model}
	\by = \bG\bs + \bz.
\end{equation}
In \eqref{equ:rece model}, $\bs$ is the real-valued transmitted symbol. Denote $\bs \triangleq \left[ s_1,s_2,\ldots,s_{2K} \right]^T \in \mathbb{S}^{2K}$, 
where $s_k \in \mathbb{S}$, 
\begin{equation*}
	\mathbb{S} \triangleq \braces{s^{\left(0\right)},s^{\left(1\right)}, \ldots,s^{\left( L-1 \right)}  }
\end{equation*}
is the alphabet for the real and imaginary components of the symmetric $\tilde{L}$-QAM modulation, and  $L = \sqrt{\tilde{L}}$.
Since $\tilde{\bz}$ is circular-symmetric complex Gaussian, it can be readily obtained that
$\bz \sim \mathcal{N}\left(\mathbf{0},\sigma_z^2\bI\right)$,
and $\sigma_z^2 = {\tilde{\sigma}_z^2}/{2}$.
Given \eqref{equ:rece model}, the \textsl{a posteriori} distribution of $\bs$ can be expressed as \cite{IGADE}
\begin{equation}\label{equ:bayesian 1}
	\begin{split}
		p\left(\bs|\by\right) &\propto p_{\textrm{pr}}\left(\bs\right)f\left(\by|\bs\right)\\
        &= \prod_{k=1}^{2K}p_{\textrm{pr},k}\left(s_k\right)f_{\textrm{G}}\left(\by;\bG\bs,\sigma_z^2\bI  \right)
	\end{split}
\end{equation}
where 
$p_{\textrm{pr}}\left(\bs\right) = \prod_{k}p_{\textrm{pr},k}\left(s_k\right)$ is the a priori probability of $\bs$,
$f\left(\by|\bs\right) = f_{\textrm{G}}\left(\by;\bG\bs,\sigma_z^2\bI  \right)$ is the PDF of $\by$ given $\bs$, 
\begin{equation*}
	p_{\textrm{pr},k}\left(s_k\right)\big|_{s_k = s^{\left(\ell\right)}} = \frac{1}{L},\ell \in \setnnega{L-1},
\end{equation*}
is the a priori probability of $s_k$.
Denote the marginals of $p\left(\bs|\by\right)$ as $\braces{p_k\left(s_k|\by\right)}_{k=1}^{2K}$.
The goal in this work is to obtain their approximations. 
Given the approximate $p_k\left(s_k|\by\right), k\in \setposi{2K}$,
we perform the maximization of the a posteriori
marginals (MPM) detection as 
\begin{equation}\label{equ:MPM dete}
	s_{k,\textrm{MPM}} = \argmax{s_k\in \mathbb{S}}p_k\left(s_k|\by\right), k\in \setposi{2K}.
\end{equation}
Consequently, the detection of the transmitted data symbol $\tilde{\bs}$ is given by
\begin{equation}\label{equ:recover of tilde_s}
	\begin{split}
			\tilde{\bs}_{\textrm{de}} =& \left[ s_{1,\textrm{MPM}}, s_{2,\textrm{MPM}}, \cdots, s_{K,\textrm{MPM}} \right]^T \\
			&+ \barjmath\left[ s_{K+1,\textrm{MPM}}, s_{K+2,\textrm{MPM}}, \cdots, s_{2K,\textrm{MPM}} \right]^T.
	\end{split}
\end{equation}

\section{GIGA}\label{sec:pre of IG}
In this section, we start by stating the signal detection
problem in perspective of information geometry. We then
express the a posteriori distribution in a factorization way
based on the grouping of the components of the received
signal. On this basis, we propose the GIGA.

\subsection{Signal Detection in Information Geometry Perspective}\label{sec:bas def}

In this subsection, we state the signal detection problem from the information geometry perspective.
We begin with the definitions of the original manifold and the objective manifold.
Define a manifold $\mathcal{S}$ as a set of probability distributions, which contains all possible probability distributions of $\bs$, i.e.,
\begin{equation}
	\mathcal{S} = \braces{p\left(\bs\right)\Big| p\left(\bs\right)>0, \bs\in \mathbb{S}^{2K},  \sum_{\bs\in\mathbb{S}^{2K}}p\left(\bs\right) = 1 }.
\end{equation}
It can be readily checked that the a prior distribution $p_{\textrm{pr}}\left(\bs\right)$ and the \textsl{a posteriori} distribution $p\left(\bs|\by\right)$ belong to $\mathcal{S}$.
We refer to $\mathcal{S}$ as the original manifold (OM).
Then, we define a sub-manifold of $\mathcal{S}$.
It is the set $\mathcal{M}_0$ of probability distributions of $\bs$, where the components of $\bs$ are independent of each other.
Define a random vector $\bt$ as 
\begin{subequations}
	\begin{equation}
		\bt \triangleq \left[ \bt_1^T,\bt_2^T,\ldots,\bt_{2K}^T \right]^T \in \bbR^{2K\left(L-1\right)},
	\end{equation}
	\begin{equation}
		\bt_k \triangleq \left[t_{k,1},t_{k,2},\ldots,t_{k,L-1}  \right]^T \in \bbR^{\left(L-1\right)},
	\end{equation}
	\begin{equation}
		t_{k,\ell} \triangleq \delta\left(s_k - s^{\left(\ell\right)}\right), \ell \in \setposi{L-1}.
	\end{equation}
\end{subequations}
We can find that the components of $\bt$ are determined by the value of $s_k$, $k\in \setposi{2K}$.
Define a vector $\bd$ as 
\begin{subequations}
	\begin{equation}
			\bd \triangleq \left[ \bd_1^T,\bd_2^T,\ldots,\bd_{2K}^T  \right]^T \in \bbR^{2K\left(L-1\right)},
	\end{equation}
	\begin{equation}
		\bd_k \triangleq \left[ d_{k,1},d_{k,2},\ldots,d_{k,L-1} \right]^T \in \bbR^{\left(L-1\right)}, 
	\end{equation}
    \begin{equation}\label{equ:d_{k,l}}
	     d_{k,\ell} =\ln \frac{p_{\textrm{pr},k}\left(s_k\right)\big|_{s_k = s^{\left(\ell\right)}}}{p_{\textrm{pr},k}\left(s_k\right)\big|_{s_k = s^{\left(0\right)}}}, \ell \in \setposi{L-1}.
    \end{equation}
\end{subequations}
We can find that $\bd$ is determined by the a priori probability of $\bs$. In fact, the marginal distribution of the a priori distribution $p_{\textrm{pr}}\left(\bs\right)$ can be expressed as \cite{IGADE}
\begin{equation*}\label{equ:p_k in exp family}
	p_{\textrm{pr},k}\left(s_k\right) = \exp\braces{\bd_k^T\bt_k - \psi\left( \bd_k \right)},
\end{equation*}
where $	\psi\left( \bd_k \right) = -\ln \left(p_{\textrm{pr},k}\left(s_k\right)\big|_{s_k = s^{\left(0\right)}}  \right) $ is the normalization factor.
Consequently, we can also obtain
\begin{equation}\label{equ:exp of a priori}
	p_\textrm{pr}\left(\bs\right) = \exp\braces{\bd^T\bt - \psi\left(\bd\right)},
\end{equation}
where $\psi\left(\bd\right) = \sum_{k=1}^{2K}\psi\left( \bd_k \right)
$ is the normalization factor.
Based on the definitions above, the sub-manifold $\mathcal{M}_0$ of $\mathcal{S}$ is defined as follows
\begin{subequations}\label{equ:M0}
	\begin{equation}
		\mathcal{M}_0 = \braces{p_0\left(\bs;\btheta_0\right)\Big|\btheta_0\in \bbR^{2K\left(L-1\right) }  },
	\end{equation}
	\begin{equation}\label{equ:defintion of p_0}
		\begin{split}
			p_0\left(\bs;\btheta_0\right) 
			&= \exp\braces{\bd^T\bt + \btheta_0^T\bt - \psi_0\left(\btheta_0\right)}\\
			&= \prod_{k=1}^{2K}p_{0,k}\left(s_k;\btheta_{0,k}\right),
		\end{split}
	\end{equation}
\end{subequations}
\begin{equation}\label{equ:marginals of p0}
	\begin{split}
		& p_{0,k}\left(s_k;\btheta_{0,k}\right) \\
		& =  \exp\braces{\bd_k^T\bt_k + \btheta_{0,k}^T\bt_k - \psi_0\left(\btheta_{0,k}\right)}\\
		& =\exp\braces{\sum_{\ell=1}^{L-1}\left( d_{k,\ell} + \theta_{0,k,\ell}\right)\delta\left(s_{k} - s^{\left(\ell\right)}\right) - \psi_0\left(\btheta_{0,k}\right) },
	\end{split}
\end{equation}
where
\begin{subequations}
	\begin{equation}\label{equ:EACS of p0}
		\btheta_0 = \left[\btheta_{0,1}^T,\btheta_{0,2}^T,\ldots,\btheta_{0,2K}^T  \right]^T \in \bbR^{2K\left(L-1\right)},
	\end{equation}
	\begin{equation}
		\btheta_{0,k} = \left[ \theta_{0,k,1},\theta_{0,k,2},\ldots,\theta_{0,k,L-1}  \right]^T\in \bbR^{\left(L-1\right)},
	\end{equation}
\end{subequations}
$p_{0,k}\left(s_k;\btheta_{0,k}\right)$ is the marginal distribution of $s_k$ given the joint distribution $p_0\left(\bs;\btheta_0\right)$,
$\psi_0\left(\btheta_{0}\right)$ is the free energy (normalization factor) of $p_0\left(\bs;\btheta_0\right)$, 
$\psi_0\left(\btheta_{0,k}\right)$ is the free energy of $p_{0,k}\left(s_k;\btheta_{0,k}\right)$, and
\begin{subequations}\label{equ:psi0}
	\begin{equation}
		\begin{split}
			\psi_0\left(\btheta_0\right) &= \sum_{k=1}^{K}\psi_0\left(\btheta_{0,k}\right)\\
			&=\ln \left( \sum_{\bs \in \mathbb{S}^{2K}}\exp\braces{ \bd^T\bt + \btheta_{0}^T\bt } \right),
		\end{split}
	\end{equation}
	\begin{equation}\label{equ:free energy of marginals of p0}
		\begin{split}
			\psi_0\left(\btheta_{0,k}\right) 
			=\ln\left( 1+ \sum_{\ell = 1}^{L-1}\exp\braces{ d_{k,\ell} + \theta_{0,k,\ell} } \right).
		\end{split}
	\end{equation}
\end{subequations}
$\btheta_0$ and $\btheta_{0,k}$ are referred to as the $e$-affine coordinate system (EACS) of $p_0\left(\bs;\btheta_0\right) $ and $p_{0,k}\left(s_k;\btheta_{0,k}\right)$, respectively.
From the definition, it can be checked that
$p_{\textrm{pr},k}\left(s_k\right) = p_{0,k}\left(s_k;\bzero\right)$, and
$p_{\textrm{pr}}\left(\bs\right) = p_0\left(\bs;\bzero\right)$.
$\mathcal{M}_0$ is referred to as the objective manifold (OBM) since it contains all the distributions of $\bs$ whose components are independent of each other, and our goal in this paper is to find a distribution in it to approximate $p\left(\bs|\by\right)$. 
In information geometry theory, this process is stated as calculating the $m$-projection of $p\left(\bs|\by\right)$ onto $\mathcal{M}_0$, i.e.,
\begin{equation}\label{equ:def of mp1}
	p_0\left(\bs;\btheta_{0}^{\star}\right) = \argmin{p_0\left(\bs;\btheta_{0}\right)\in \mathcal{M}_0} \Dkl{p\left(\bs|\by\right)}{p_0\left(\bs;\btheta_{0}\right)},
\end{equation}
where $\Dkl{}{}$ is the Kullback-Leibler (K-L) divergence, and
\begin{equation*}
	\Dkl{p\left(\bs|\by\right)}{p_0\left(\bs;\btheta_{0}\right)} = \sum_{\bs\in \mathbb{S}^{2K}}p\left(\bs|\by\right)\ln\left(\frac{p\left(\bs|\by\right)}{p_0\left(\bs;\btheta_{0}\right)}\right).
\end{equation*}
$p_0\left(\bs;\btheta_{0}^{\star}\right)$ can be interpreted as the distribution in $\mathcal{M}_0$ that is closest to $p\left(\bs|\by\right)$, where the distance between the two distributions is defined as the K-L divergence.
Given $p_0\left(\bs;\btheta_{0}^{\star}\right)$, its marginals can be directly obtained since the components of $\bs$ are independent.
On the other hand, the calculation of
the direct $m$-projection may be unacceptable since it can be too complicated.
To solve this problem, we find another distribution in the OBM $\mathcal{M}_0$ to approximate $p\left(\bs|\by\right)$ by grouping the components of the received signal $\by$ and defining an extra type of manifolds.

\subsection{Factorization of $p\left(\bs|\by\right)$ by Grouping the Components of the Received Signal}

In this subsection, we factorize the \textsl{a posteriori} distribution $p\left(\bs|\by\right)$  by the grouping the components of the received signal $\by$.
We first present the way of grouping.
Define the set of indexes of all components of $\by$ as $\mathcal{N}_{\textrm{o}} \triangleq \braces{1,2,\ldots,2N_r}$.
We uniformly divide $\mathcal{N}_\textrm{o}$ into $U$ subsets, where $U$ is a factor of $2N_r$. 
Then, the number of the elements in each subset is 
\begin{equation*}
	N_u = \frac{2N_r}{U},
\end{equation*}
Denote each subset as $\mathcal{N}_u$, $u \in \setposi{U}$ and we have
\begin{equation}
	\mathcal{N}_u = \braces{ \left(u-1\right)N_u +1, \left(u-1\right)N_u +2, \cdots, uN_u  }.
\end{equation}

%
%
%

According to the subsets $\braces{\mathcal{N}_u}_{u=1}^{U}$, we define $U$ sub-vectors of $\by$, where the $u$-th of them  only contains the components of $\by$ indexed in $\mathcal{N}_u$, i.e., 
\begin{equation}\label{equ:definition of yu}
	\by_{u} = \left[ \by \right]_{\mathcal{N}_u} \in \bbR^{N_u}.
\end{equation}
Given $\bs$, the PDF of $\by_{u}$ is Gaussian, and we have
\begin{equation}
	\begin{split}
		f\left(\by_{u}|\bs  \right) = f_{\textrm{G}}\left(\by_{u};\bG_u\bs,\sigma_z^2\bI  \right) 
	\end{split}
\end{equation}
where
\begin{equation}\label{equ:G_u}
\bG_u = \left[ \bG \right]_{\mathcal{N}_u,:} \in \bbR^{N_u \times 2K}
\end{equation} 
is a sub-matrix of $\bG$ in \eqref{equ:rece model}.
Since given $\bs$, all components of $\by$ in \eqref{equ:rece model} are independent with each other, thus $\braces{\by_u}_{u=1}^U$ are also independent, and we can readily obtain that $f\left(\by|\bs\right) = \prod_{u=1}^{U}f\left(\by_{u}|\bs\right)$.
On this basis, the \textsl{a posteriori} distribution $p\left(\bs|\by\right)$ can be factorized as 
\begin{equation}\label{equ:fac 1}
	p\left(\bs|\by\right) \propto \prod_{k=1}^{2K}p_{\textrm{pr},k}\left(s_k\right)\prod_{u=1}^{U}f\left(\by_{u}|\bs\right).
\end{equation}

\subsection{GIGA}

According to \eqref{equ:fac 1}, the \textsl{a posteriori} probability distribution $p\left(\bs|\by\right)$ can be further expressed as
\begin{equation}\label{equ:post distribution}
	p\left(\bs|\by\right) 
	= \exp\braces{  \bd^T\bt + \sum_{u=1}^{U}c_u\left( \bs,\by_{u} \right) - \psi_q },
\end{equation}
where
\begin{equation}\label{equ:cu}
	\begin{split}
		c_u\left(\bs,\by_{u}\right) 
		= -\frac{1}{2\sigma_z^2}\norm{\by_{u} - \bG_u\bs }^2,
	\end{split}
\end{equation}
$\psi_q$ is the normalization factor, and
\begin{equation}
	\psi_q = \ln \left( \sum_{\bs \in \mathbb{S}^{2K}}\exp\braces{ \bd^T\bt +  \sum_{u=1}^{U}c_u\left( \bs,\by_u \right) } \right).
\end{equation}
In \eqref{equ:post distribution}, $\bd^T\bt$ only contains a linear combination of single independent random variables $s_k$, and $\braces{c_u\left( \bs,\by_{u} \right)}_{u=1}^{U}$ contain all the interactions (cross-terms) between the random variables, i.e., $s_ks_{k'}$, $k \neq k'$.
If we can approximate the sum $\sum_{u=1}^{U}c_u\left(\bs,\by_{u}\right)$ as $\btheta_{0}^T\bt$, then we have
\begin{equation}
	p\left( \bs|\by \right) \approx p_0\left(\bs;\btheta_{0}\right) = \exp\braces{\left(\bd + \btheta_{0}\right)^T\bt - \psi_0},
\end{equation}
where $\psi_0$ is the free energy, which is simple to use.
In this work, we obtain the approximations of $\braces{c_u\left( \bs,\by_{u} \right)}_{u=1}^{U}$ through iterative $m$-projections.
Let us first define a class of sub-manifolds of $\mathcal{S}$.
Given the number $U$ of subsets, we define $U$ sub-manifolds of $\mathcal{S}$, where the $u$-th of them is given by
\begin{subequations}\label{equ:AMs}
	\begin{equation}
		\mathcal{M}_u = \braces{ p_u\left(\bs;\btheta_u\right) \Big| \btheta_u \in \bbR^{2K\left(L-1\right)}  },
	\end{equation}
	\begin{equation}\label{equ:pn}
		p_u\left(\bs;\btheta_u\right) = \exp\braces{ \bd^T\bt + \btheta_u^T\bt + c_u\left(\bs,\by_{u}\right) - \psi_u\left(\btheta_u\right)  },
	\end{equation}
\end{subequations}
where 
\begin{subequations}\label{equ:EACS of pn}
	\begin{equation}
		\btheta_u = \left[\btheta_{u,1}^T,\btheta_{u,2}^T,\ldots,\btheta_{u,2K}^T  \right]^T \in \bbR^{2K\left(L-1\right)},
	\end{equation}
	\begin{equation}
		\btheta_{u,k} = \left[ \theta_{u,k,1},\theta_{u,k,2},\ldots,\theta_{u,k,L-1}  \right]^T\in \bbR^{\left(L-1\right)},
	\end{equation}
\end{subequations}
$c_u\left(\bs,\by_{u}\right)$ is given by \eqref{equ:cu},
and the free energy $\psi_u$ is given by 
\begin{equation}
	\psi_u\left(\btheta_u\right) = \ln \left( \sum_{\bs \in \mathbb{S}^{2K}}\exp\braces{ \bd^T\bt + \btheta_u^T\bt + c_u\left( \bs,\by_{u} \right) } \right).
\end{equation}
$\braces{\mathcal{M}_u}_{u=1}^U$ are referred to as the auxiliary manifolds (AMs), and $\btheta_{u}$ is referred to as the EACS of $p_u\left(\bs;\btheta_n\right)$.
Note that AMs will vary with the number $U$ of subsets  since the definitions of $\by_{u}$ and $\bG_u$ will vary with $U$.
Compared to $p\left(\bs|\by\right)$ in \eqref{equ:post distribution}, $p_u\left(\bs;\btheta_u\right)$ reserves only one interaction term $c_u\left( \bs,\by_{u} \right)$, while all others, i.e., $\sum_{u' \neq u}c_u\left(\bs,\by_{u}\right)$, are replaced with $\btheta_{u}^T\bs$.

\begin{figure}[t]
	\centering
	\includegraphics[width=0.32\textwidth]{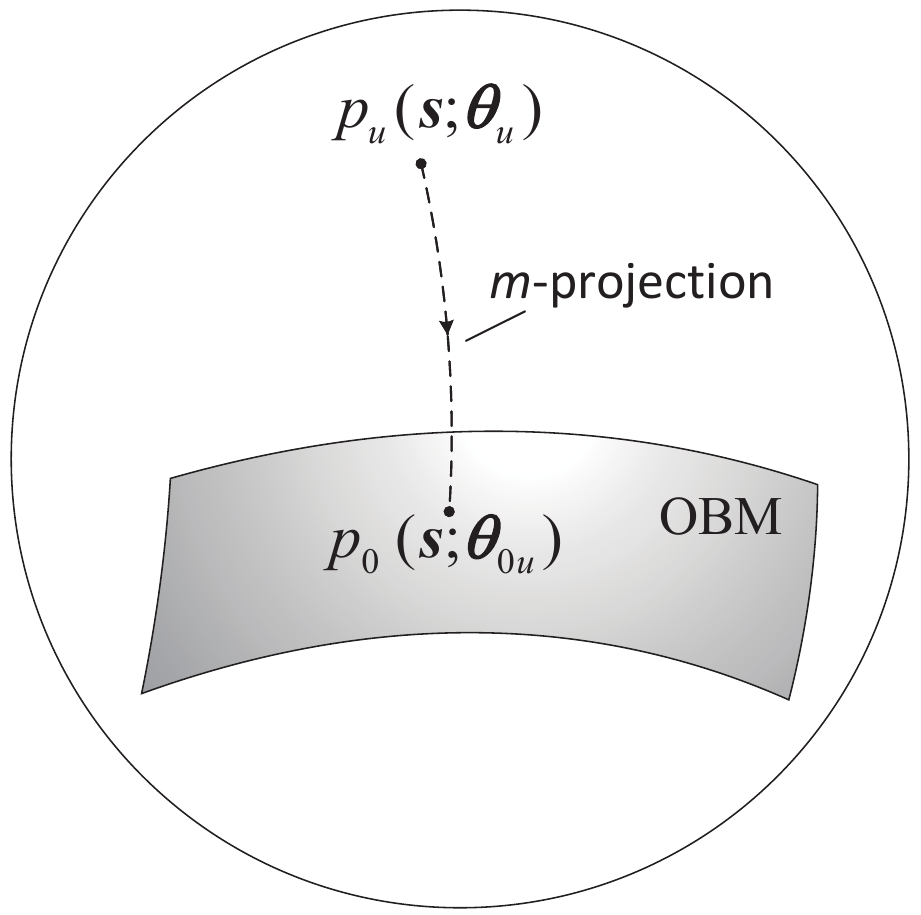}
	\caption{ Illustration of the $m$-projection of $p_u\left(\bs;\btheta_{u}\right)$ onto the OBM.}
	\label{fig:m-p}
\end{figure}

$p_u\left(\bs;\btheta_u\right)$ of $\mathcal{M}_u$ is the key to obtain the approximation of single interaction term $c_u\left(\bs,\by_{u}\right)$.
Given $p_u\left(\bs;\btheta_u\right)$, we can obtain an approximation of $c_u\left(\bs,\by_{u}\right)$ through its $m$-projection on the OBM, which is illustrated in Fig. \ref{fig:m-p}.
Denote this $m$-projection as $p_0\left(\bs;\btheta_{0u}\right)$, we have
\begin{equation*}
	p_0\left(\bs;\btheta_{0u}\right) = \argmin{p_0\left(\bs;\btheta_{0}\right)\in \mathcal{M}_0} \Dkl{p_u\left(\bs;\btheta_{u}\right)}{p_0\left(\bs;\btheta_{0}\right)},
\end{equation*}
which can be expressed in a more explicit way as
\begin{equation}\label{equ:theta_{0n} in preli}
	\btheta_{0u} = \argmin{\btheta_{0}\in \bbR^{2K\left(L-1\right)}}\Dkl{p_u\left(\bs;\btheta_{u}\right)}{p_0\left(\bs;\btheta_{0}\right)}.
\end{equation}
In  Sec. \ref{sec:GIGA-SD}, we will present a detailed calculating procedure.
Here, we focus on the next steps.
Suppose that $\btheta_{0u}$ is obtained,
we re-express the $m$-projection $p_0\left(\bs;\btheta_{0u}\right)$ in the following way:
\begin{equation}\label{equ:mp preli}
	\begin{split}
		p_0\left(\bs;\btheta_{0u}\right) &= \exp\braces{\left( \bd+\btheta_{0u} \right)^T\bt - \psi_0\left(\btheta_{0u}\right)}\\
		&= \exp\braces{\left( \bd+\btheta_{u}+\bxi_{u} \right)^T\bt- \psi_0\left(\btheta_{0u}\right)},
	\end{split}
\end{equation}
which implies that $\btheta_{0u} = \btheta_{u}+\bxi_{u}$.
In \eqref{equ:mp preli}, we consider the EACS $\btheta_{0u}$ of $p_0\left(\bs;\btheta_{0u}\right)$  as a sum of the EACS $\btheta_{u}$ of $p_n\left(\bs;\btheta_{u}\right)$ and an extra item $\bxi_{u}$.
Comparing $p_u\left(\bs;\btheta_{u}\right)$ defined in \eqref{equ:pn}  and \eqref{equ:mp preli},  we can find that $c_u\left(\bs,\by_{u}\right)$ in $p_u\left(\bs;\btheta_{u}\right)$ is replaced by $\bxi_{u}^T\bt$ in $p_0\left(\bs;\btheta_{0u}\right)$.
Consequently, we consider the approximation of $c_u\left(\bs,\by_{u}\right)$ as $\bxi_{u}^T\bt$, where 
\begin{equation}\label{equ:xi_n in preli}
	\bxi_{u} = \btheta_{0u} - \btheta_{u}, u\in \setposi{U}.
\end{equation}
Then,  $p_0\left(\bs;\btheta_{0}\right)$ with $\btheta_{0} = \sum_{u=1}^{U}\bxi_{u}$ is considered as the approximation of the \textsl{a posteriori} distribution $p\left(\bs|\by\right)$ since each $c_u\left(\bs,\by_{u}\right)$ is approximated as $\bxi_{u}^T\bt$ and $\btheta_{0}^T\bt$ is regarded as an approximation of $\sum_{u}c_u\left(\bs,\by_{u}\right)$.

Now, let us discuss the iteration,
which is similar to that for IGA in \cite{IGADE}.
At the beginning, the EACSs are initialized as $ \braces{\btheta_{u}\left(0\right)}_{u=0}^{U}$.
Then, given the EACSs $\braces{\btheta_{u}\left(t\right)}_{u=0}^{U}$ at the $t$-th iteration, calculate  $\braces{\btheta_{0u}\left(t\right)}_{u=1}^U$ and $\braces{\bxi_{u}\left(t\right)}_{u=1}^U$ as \eqref{equ:theta_{0n} in preli} and \eqref{equ:xi_n in preli}, respectively.
Next, the EACS of $p_u\left(\bs;\btheta_{u}\left(t\right)\right), u\in \setposi{U}$ is updated as
\begin{equation*}
	\btheta_{u}\left(t+1\right) = \sum_{u' = 1, u' \neq u}^{U}\bxi_{u'}\left(t\right),
\end{equation*}
since  $\btheta_{u}^T\left(t+1\right)\bt$ replaces 
the sum $\sum_{u' \neq u}^Uc_{u'}\left(\bs,\by_{u'} \right)$ 
in $p_u\left(\bs;\btheta_{u}\left(t+1\right)\right)$ and each $c_u\left( \bs,\by_{u} \right)$ is approximated as $\bxi_{u}^T\left(t\right)\bt$ at the $t$-th iteration.
We update the EACS of $p_0\left(\bs;\btheta_{0}\left(t\right)\right)$ as $\btheta_{0}\left(t+1\right) = \sum_{u=1}^{U}\bxi_{u}\left(t\right)$.
In practice, the following damped updating way is typically employed to ensure the convergence: given a damping factor $0<\alpha\le 1$, update the EACSs as
\begin{subequations}\label{equ:update of NPs in GIGA}
	\begin{equation}
		\btheta_{u}\left(t+1\right) = \alpha \sum_{u' = 1, u'\neq u}^{U}\bxi_{u'}\left(t\right) + \left(1-\alpha\right)\btheta_{u}\left(t\right), u\in \setposi{U},
	\end{equation}
	\begin{equation}
		\btheta_{0}\left(t+1\right) = \alpha\sum_{u=1}^{U}\bxi_u\left(t\right)+\left(1-\alpha\right)\btheta_{0}\left(t\right).
	\end{equation}
\end{subequations}
Then, repeat the above process until convergence.
We refer to the above process as GIGA, and we summarize it in Algorithm \ref{Alg:IGA}.
Note that the procedure for calculating the $m$-projection in Step $1$ will be discussed in the next section.
\begin{algorithm}[t]
	\SetAlgoNoLine 
	\caption{GIGA}
	\label{Alg:IGA}
	
	\KwIn{The a priori probability $p_{\textrm{pr},k}\left(s_k\right), k\in \setposi{2K}$, the received signal $\by$, the channel matrix $\bG$, the alphabet $\mathbb{S} = \braces{s^{\left(0\right)},s^{\left(1\right)}, \ldots,s^{\left( L-1 \right)}}$, the number $U$ of subsets,
		the noise power ${\sigma}_z^2$ and the maximal iteration number $t_{\mathrm{max}}$.}
	
	\textbf{Initialization:} Set $t=0$. Set damping $\alpha$, where $0< \alpha\le 1$.
	Calculate $\by_{u}$ and $\bG_u$, $u \in \setposi{U}$, as \eqref{equ:definition of yu} and \eqref{equ:G_u}, respectively.
	Initialize the EACSs $\braces{\btheta_{u}}_{u=0}^U$ defined in \eqref{equ:EACS of p0} and \eqref{equ:EACS of pn}. Zeros are sufficient for the initializations. 
	Calculate the NP $d_{k,\ell}$, $k\in \setposi{2K}$, $\ell \in \setposi{L-1}$, as \eqref{equ:d_{k,l}}; 
	
	\Repeat{\rm{Convergence or $t > t_{\mathrm{max}}$}}{
		1. Calculate  the EACS $\btheta_{0u}$, $u \in \setposi{U}$, of the $m$-porjection in \eqref{equ:theta_{0n} in preli};\\ 
		2. Calculate $\bxi_u{\left(t\right)}, n\in \setposi{2N_r}$, as \eqref{equ:xi_n in preli};\\
		3. Update the EACSs as
		\eqref{equ:update of NPs in GIGA};\\
		4. $t = t+1$;}
	
	\KwOut{\rm{The approximate marginals, $\braces{p_k\left( s_k|\by\right)}_{k=1}^{2K}$, are given by  $\braces{p_{0,k}\left(s_k;\btheta_{0,k}\right)}_{k=1}^{2K}$, which is defined in \eqref{equ:probability of marginals of p0}. Perform the MPM detection as \eqref{equ:MPM dete}, and $\tilde{\bs}$ is recovered as \eqref{equ:recover of tilde_s}.}}
\end{algorithm}

\section{Calculation of the $m$-projection from  \texorpdfstring{$p_u\left(\bs;\btheta_u\right)$}{} onto the OBM}\label{sec:GIGA-SD}
In this section, we present the calculation of the $m$-projection from $p_u\left(\bs;\btheta_u\right)$ onto the OBM.
We first give its direct calculation.
Then, based on the Berry-Esseen theorem, we propose an approximate calculation of the $m$-projection. 
The efficient implementation of the approximate calculation is also discussed.
At last, we analyze the complexities of two types of calculations of $m$-projections.

\subsection{Direct Calculation}
The direct calculation of the $m$-projection in \eqref{equ:theta_{0n} in preli} is related to the marginal distributions of $p_u\left(\bs;\btheta_{u}\right)$.
To express the marginals of $p_u\left(\bs;\btheta_{u}\right)$, let us first define $2K$ discrete random vectors with the same dimension, where the $k$-th vector is denoted as $\bs_{\setminus k}$ and $\bs_{\setminus k}$ is of $2K-1$ dimension. 
$\bs_{\setminus k}$ is obtained by removing the $k$-th component, i.e., $s_k$, in $\bs$.
Based on the definition, we have $\bs_{\setminus k}\in \mathbb{S}^{2K-1}, k\in \setposi{2K}$.
Given $p_u\left(\bs;\btheta_{u}\right), u\in\setposi{U}$, we denote its marginal probability distribution of $s_k$ as $p_{u,k}\left(s_k;\btheta_{u}\right)$, and we have 
\begin{equation}\label{equ:marginal of pu}
	\begin{split}
	   	&p_{u,k}\left(s_k;\btheta_{u}\right) \\
	   	=&\sum_{s_1\in \mathbb{S}}\cdots \sum_{s_{k-1}\in \mathbb{S}}\sum_{s_{k+1}\in \mathbb{S}}\cdots\sum_{s_{2K}\in \mathbb{S}} 	p_u\left(\bs;\btheta_u\right)\\
	   	=& \sum_{\bs_{\setminus k} \in \mathbb{S}^{2K-1}}p_u\left(\bs;\btheta_u\right) .	   
	\end{split}
\end{equation}
Further, denote the EACS of the $m$-projection $p_0\left(\bs;\btheta_{0u}\right)$ as
\begin{subequations}\label{equ:def of mp}
	\begin{equation}
		\btheta_{0u} = \left[ \btheta_{0u,1}^T,\btheta^T_{0u,2},\ldots,\btheta^T_{0u,2K} \right]^T \in \bbR^{2K\left(L-1\right)},	
	\end{equation}
	\begin{equation}
		\btheta_{0u,k} = \left[ \theta_{0u,k,1}, \theta_{0u,k,2}, \ldots, \theta_{0u,k,L-1} \right]^T \in \bbR^{\left(L-1\right)}.
	\end{equation}
\end{subequations}
Based on \eqref{equ:M0}, denote the marginals of $p_0\left(\bs;\btheta_{0u}\right)$ as $p_0\left(s_k;\btheta_{0u,k}\right)$, $k\in \setposi{2K}$.
Then, according to \cite[Theorem 1]{IGADE}, $\btheta_{0u}$ exists and is unique, and we have
\begin{equation}\label{equ:mp1}
	p_0\left(s_k;\btheta_{0u,k}\right) = p_u\left(s_k;\btheta_{u}\right), s_k \in \mathbb{S},
	k \in \setposi{2K}.
\end{equation}
We can find that the marginals of $p_u\left(\bs;\btheta_{u}\right)$ and its $m$-projection are the same.

We next discuss how to obtain the value of the EACS $\btheta_{0u}$ from \eqref{equ:mp1}.
This is related to the property of $p_0$ of OBM.
Given any $p_0\left( \bs;\btheta_{0} \right)$ in the OBM and its marginals $p_{0,k}\left(s_k;\btheta_{0,k}\right)$, we have \cite{IGADE}
\begin{subequations}\label{equ:probability of marginals of p0}
	\begin{equation}\label{equ:probability of mariginals of p0 on s^0}
		p_{0,k}\left(s_k;\btheta_{0,k}\right)\Big|_{s_{k} = s^{\left(0\right)}} = \frac{1}{1+ \sum_{\ell = 1}^{L-1}\exp\braces{ d_{k,\ell} + \theta_{0,k,\ell} }},
	\end{equation}
	\begin{equation}\label{equ:probability of mariginals of p0 on s^ell}
		p_{0,k}\left(s_k;\btheta_{0,k}\right)\Big|_{s_{k} = s^{\left(\ell\right)}} = \frac{\exp\braces{ d_{k,\ell} + \theta_{0,k,\ell} }}{1+ \sum_{\ell = 1}^{L-1}\exp\braces{ d_{k,\ell} + \theta_{0,k,\ell} }},
	\end{equation}
\end{subequations}
where $\ell \in \setposi{L-1}$ in \eqref{equ:probability of mariginals of p0 on s^ell}.
On the contrary, given the probability in \eqref{equ:probability of marginals of p0}, the EACS of $p_{0,k}\left(s_k;\btheta_{0,k}\right), k\in \setposi{2K}$ can be expressed as 
\begin{equation}\label{equ:relation}
	\theta_{0,k,\ell} = \ln\frac{p_{0,k}\left(s_k;\btheta_{0,k}\right)\Big|_{s_{k} = s^{\left(\ell\right)}}}{p_{0,k}\left(s_k;\btheta_{0,k}\right)\Big|_{s_{k} = s^{\left(0\right)}}} - d_{k,\ell}, \ell\in \setposi{L-1}.
\end{equation}
Since the $m$-projection $p_0\left(\bs;\btheta_{0u}\right)$ belongs to the OBM,  $\theta_{0u,k,\ell}$ in \eqref{equ:mp1} can be expressed as 
\begin{equation}\label{equ:mp solution}
	\theta_{0u,k,\ell} = \ln\frac{p_{u,k}\left(s_k;\btheta_{u}\right)\big|_{s_k = s^{\left(\ell\right)}}}{p_{u,k}\left(s_k;\btheta_{u}\right)\big|_{s_k = s^{\left(0\right)}}} - d_{k,\ell},
\end{equation}
which shows that the EACS of the $m$-projection $p_0\left(\bs;\btheta_{0u}\right)$ is determined by the marginal probability of $p_u\left(\bs;\btheta_{u}\right)$.
On the other hand, the closed-form solution of  $p_{u,k}\left(s_k;\btheta_{u}\right)$ can be hard to obtain. 
From \eqref{equ:marginal of pu} we can find that its calculation is of exponential-complexity.
When the number of users and the modulation order are large, the computational complexity of \eqref{equ:marginal of pu} is unaffordable.
Inspired by the Berry-Esseen theorem, we solve this problem by computing an approximation of the marginal $p_{u,k}\left(s_k;\btheta_{u}\right)$, $u \in \setposi{U}$, $k \in \setposi{2K}$, which will be discussed in the next subsection.

\subsection{Approximate Calculation}
As mentioned above, our focus now is to calculate the approximate marginals of $p_u\left(\bs;\btheta_{u}\right)$.
To do so, we first express $p_{u,k}\left(s_k;\btheta_{u}\right)$  as follows:
\begin{align}\label{equ:marginals of pn}
		p_{u,k}\left(s_k;\btheta_u\right) &=\!\!\! \sum_{\bs_{\setminus k} \in \mathbb{S}^{2K-1}}\!\!\!\exp\braces{ \left(\bd +\btheta_u  \right)^T \bt + c_u\left(\bs,\by_{u}\right) - \psi_u }\nonumber\\
		&\overset{\left(\textrm{a}\right) }{\propto}\lambda_{u,k}\left(s_k\right)
		\kappa_{u,k}\left( s_k,\by_{u} \right), 
\end{align}
where $u \in \setposi{U}, k\in \setposi{2K},$
$s_k \in \mathbb{S}$,
$\left(\textrm{a}\right)$ is obtained by removing the constant independent with $s_k$,
and
\begin{subequations}
	\begin{equation}
		\begin{split}
			\lambda_{u,k}\left(s_k\right) &\triangleq \exp\braces{\left( \bd_k+\btheta_{u,k} \right)^T\bt_k }\\
			&= \exp\braces{\sum_{\ell=1}^{L-1} \left(d_{k,\ell} + \theta_{u,k,\ell} \right) \delta\left( s_k - s^{\left(\ell\right)} \right)},
		\end{split}
	\end{equation}
	\begin{align}\label{equ:q(sk,yn)}
		&\kappa_{u,k}\left(s_k,\by_{u}\right)\\
		\triangleq &\sum_{\bs_{\setminus k} \in \mathbb{S}^{2K-1}}\exp\Big\{ \sum_{\substack{k'=1, k'\neq k}}^{2K}\left(\bd_{k'}+ \btheta_{u,k'}  \right)^T\bt_{k'} + c_u\left( \bs,\by_{u} \right)  \Big\}.	\nonumber
	\end{align}
\end{subequations}
The reason why we explicitly parameterize $\kappa_{u,k}\left(s_k,\by_{u}\right)$ by both $s_k$ and $\by_{u}$ is that $\by_{u}$ will play an important role when computing an approximation of $\kappa_{u,k}\left(s_k,\by_{u}\right)$. 
In \eqref{equ:marginals of pn}, the value of $\lambda_{u,k}\left(s_k\right)$ can be calculated directly, since we have 
\begin{equation}
	\lambda_{u,k}\left(s_k\right) =
	\begin{cases}
		1, &s_k = s^{\left(0\right)},\\
		\exp\braces{d_{k,\ell} + \theta_{u,k,\ell}}, &s_k = s^{\left(\ell\right)},\ell \in \setposi{L-1}.
	\end{cases}
\end{equation}
Under these circumstances, calculating the approximation of $\kappa_{u,k}\left(s_k,\by_{u}\right)$, $s_k \in \mathbb{S}$, becomes the critical issue.
Once it is obtained, the approximate value of $p_{u,k}\left(s_k;\btheta_{u}\right)$, $s_k \in \mathbb{S}$, can be acquired.
Also, as a note, the proportion in \eqref{equ:marginals of pn} does not influence the computation of $p_{u,k}\left(s_k;\btheta_{u}\right)$ because
the constant corresponding to the proportion is independent with $s_k$. 
Therefore, the value of $p_{u,k}\left(s_k;\btheta_{u}\right)$ can be always obtained through $\sum_{s_k \in \mathbb{S}}p_{u,k}\left(s_k;\btheta_{u}\right) = 1$.
We will not repeat this issue when similar situations arise in the next.

For now, our attention shifts to the value of $\kappa_{u,k}\left(\by_{u},s_k\right), s_k\in \mathbb{S}$.
According to its definition, $\kappa_{u,k}\left(s_k,\by_{u}\right)$ can be further expressed as
\begin{align}\label{equ:q(sk,yn)2}
		& \ \ \ \ \kappa_{u,k}\left(s_k,\by_{u}\right)\nonumber \\
		&= \sum_{\bs_{\setminus k} \in \mathbb{S}^{2K-1}}\Big( \prod_{k'=1,k'\neq k}^{2K}\exp\braces{ \left( \bd_{k'}+ \btheta_{u,k'} \right)^T \bt_{k'} }\\
		 & \ \ \times\exp\braces{ -\frac{1}{2}\left( \by_{u} - \bG_u\bs \right)^T\left(\sigma_z^2\bI\right)^{-1}\left( \by_{u} - \bG_u\bs \right) }  \Big)\nonumber \\
		 &\overset{\left(\textrm{a}\right)}{\propto} \!\! \sum_{\bs_{\setminus k} \in \mathbb{S}^{2K-1}} \Big( \prod_{k'=1, k'\neq k}^{2K}p_{0,k'}\left( s_{k'};\btheta_{u,k'} \right) f_{\textrm{G}}\left( \by_{u};\bG_u\bs,\sigma_z^2\bI \right) \Big),\nonumber
\end{align}
where $u \in \setposi{U}$, $k \in \setposi{2K}$, and
$\left(\textrm{a}\right)$ comes from adding a constant independent with $s_k$ and $\by_{u}$.
Next we construct $2KU$ random vectors whose PDFs are in the same form as the last line of \eqref{equ:q(sk,yn)2}. 
The introduction of these random vectors is the key to compute the approximations of $\kappa_{u,k}\left(s_k,\by_{u}\right)$, $u \in \setposi{U}$, $k \in \setposi{2K}$.
Define $U\times 2K$ random vectors 
\begin{equation*}
	\braces{\overline{\by}_{u,k}|u\in \setposi{U}, k\in \setposi{2K}}.
\end{equation*} 
The $\left(u,k\right)$-th vector is given by
\begin{equation}\label{equ:Yn}
	\begin{split}
		\overline{\by}_{u,k} &= 
		\bg_{u,k}s_k + \sum_{k'=1, k'\neq k}^{2K}\bg_{u,k'}s_{k'} + \bw \\
		&= \sum_{k'=1, k'\neq k}^{2K}\bg_{u,k'}s_{k'} + \bw'_{u,k},
	\end{split}
\end{equation}
where $u\in \setposi{U}, k \in \setposi{2K}$,
$\bg_{u,k} \in \bbR^{N_u}$ is the $k$-th column of $\bG_u$ in \eqref{equ:G_u},  i.e.,
\begin{equation*}
	\bG_u = \left[ \bg_{u,1}, \bg_{u,2}, \cdots, \bg_{u,2K}\right],
\end{equation*}
$\bg_{u,k}$ is considered as a determinate and known vector, 
$s_k$ is considered as a determinate and known constant,
\begin{equation*}
	\braces{s_{k'}}_{k'=1, k'\neq k}^{2K}
\end{equation*}
are considered as $2K-1$ independent discrete random variables,
their probability distributions are $\braces{p_{0,k'}\left(s_{k'};\btheta_{u,k'}\right)}_{k'\neq k}$,
$\bw \sim \mathcal{N}\left(\mathbf{0},\sigma_z^2\bI\right)$ is a Gaussian random vector of $N_u$ dimension independent with $\braces{s_{k'}}_{k'\neq k}$, 
and $\bw'_{u,k} = \bw + \bg_{u,k}s_k \sim \mathcal{N}\left( \bg_{u,k}s_k,\sigma_z^2\bI \right)$ is also a Gaussian random vector independent with $\braces{s_{k'}}_{k'\neq k}$.
The joint probability distribution of  $\braces{s_{k'}}_{k'\neq k}$ in \eqref{equ:Yn} is  given by
\begin{equation*}
p\left( \bs_{\setminus k} \right) = \prod_{k'=1, k'\neq k}^{2K}p_{0,k'}\left(s_{k'};\btheta_{u,k'} \right).	
\end{equation*}
Then, the PDF of $\overline{\by}_{u,k}$ can be expressed as \cite[Sec. 6.1.2]{pishro2016introduction} 
\begin{align}\label{equ:PDF of yn}
	   	&f\left(\overline{\by}_{u,k}\right)\nonumber\\
	   	 = & \sum_{\bs_{\setminus k} \in \mathbb{S}^{2K-1}}\left(\! p\left(\bs_{\setminus k}\right)f_{\textrm{G}}\left(\!\overline{\by}_{u,k}\!-\!\sum_{k'\neq k}\bg_{u,k'}s_{k'} ;\bg_{u,k}s_k,\sigma_z^2\bI  \!\right) \!\right) \nonumber\\
	   	=& \sum_{\bs_{\setminus k} \in \mathbb{S}^{2K-1}}\left(p\left(\bs_{\setminus k}\right)f_{\textrm{G}}\left( \overline{\by}_{u,k};\bG_u\bs,\sigma_z^2\bI \right)  \right).  	
\end{align}
It is a direct result that 
the PDF of $\overline{\by}_{u,k}$ will be equivalent to the final line of \eqref{equ:q(sk,yn)2}  when we substitute $\overline{\by}_{u,k}$ with $\by_{u}$.
Consequently, we can obtain
\begin{equation}\label{equ:kappa app 1}
	 \kappa_{u,k}\left(s_k,\by_{u}\right) \propto f\left(\overline{\by}_{u,k}  \right)\Big|_{\overline{\by}_{u,k} = \by_u}, s_k\in \mathbb{S}.
\end{equation}

$\overline{\by}_{u,k}$ in \eqref{equ:Yn} is a hybrid random vector, which is the sum of $2K-1$ discrete random vectors and one Gaussian random vector. 
The closed-form solution of its PDF is difficult to obtain, and as can be seen from \eqref{equ:PDF of yn}, its computational complexity is  exponential.
We note that $\overline{\by}_{u,k}$ is obtained by summing multiple mutually independent random vectors.
This is somewhat similar to the situation described by the central limit theorem, with the difference that we are dealing with the summation of multiple random vectors.
In this case, the classic central limit theorem can not be applied to obtain an approximation of the probability distribution of $\overline{\by}_{u,k}$.
Fortunately, Berry–Esseen theorem can help us to obtain such an approximation.
We first present the Berry–Esseen theorem. 
\begin{lemma}[Berry–Esseen theorem]\label{lemma:CLT}
	Given $N$ independent random vectors $\braces{\bx_n}_{n=1}^N$, where $\bx_n \in \bbR^{d}$, and $d \ge 1$ is finite.
	Each $\bx_n$ has finite mean $\bmu_n \in \bbR^{d}$ and finite positive-definite covariance matrix $\bSigma_n \in \bbR^{d\times d}$.
	Denote the summation of $\braces{\bx_n}_{n=1}^N$ as $\bx_s \triangleq \sum_{n=1}^{N}\bx_n$. Its mean and covariance matrix are denoted as $\bmu_s \triangleq \sum_{n=1}^{N}\bmu_n$ and $\bSigma_s \triangleq \sum_{n=1}^{N}\bSigma_n$, respectively.
    If the following condition
	\begin{equation*}\label{equ:Lyapunov condition}
		\lim\limits_{N\to\infty}\sum_{n=1}^{N}\Exp\braces{ \norm{ \bSigma_s^{-\frac{1}{2}} \left(\bx_n - \bmu_n\right) }^3 }\\
		    = 0		    
	\end{equation*} 
    holds.
    Then,  $\bx_s$ converges in distribution to a real Gaussian random vector $\tilde{\bx}_s$, as $N$ tends to infinity, i.e.,
\begin{equation*}
	\bx_s\to\tilde{\bx}_s  \sim \mathcal{N}\left(\bmu_s,\bSigma_s\right).
\end{equation*}
\end{lemma}

Inspired by the Berry-Esseen theorem, we approximate the PDF of $\overline{\by}_{u,k}$ as a Gaussian PDF 
\begin{equation}\label{equ:Gaussian PDF}
	f_{\textrm{G}}\left( \overline{\by}_{u,k}; \Exp\braces{\overline{\by}_{u,k}},\mathbb{V}\braces{\overline{\by}_{u,k}}\right),
\end{equation}
where $\Exp\braces{\overline{\by}_{u,k}}$ and $\mathbb{V}\braces{\overline{\by}_{u,k}} $ are the mean and covariance of $\overline{\by}_{u,k}$, respectively. 
In comparison to the original PDF $f\left(\overline{\by}_{u,k}\right)$, \eqref{equ:Gaussian PDF} has an explicit expression and it contains only linear operations.
We next calculate the mean and covariance of $\overline{\by}_{u,k}$.
To do so, we first calculate the mean and variance of $s_{k'}, k' \neq k$, in \eqref{equ:Yn}. 
The probability distribution of $s_{k'}, k' \neq k$, is $p_{0,k'}\left( s_{k'};\btheta_{u,k'} \right)$.
According to \eqref{equ:probability of marginals of p0}, the mean and the variance of $s_{k'}, k' \neq k$, are given by
\begin{subequations}\label{equ:mean and variance of marginals of p0}
	\begin{equation}\label{equ:mu_{n,k}}
		\begin{split}
		\mu_{u,k'} &= \sum_{s_{k'} \in \mathbb{S}}s_{k'}p_{0,k'}\left( s_{k'};\btheta_{u,k'} \right)\\
		&=\frac{ s^{\left(0\right)} + \sum_{\ell=1}^{L-1}s^{\left(\ell\right)} \exp\braces{d_{k',\ell} + \theta_{u,k',\ell}} }{1+\sum_{\ell=1}^{L-1}\exp\braces{ d_{k',\ell} + \theta_{u,k',\ell} }  },
		\end{split}
	\end{equation}
\begin{equation}\label{equ:v_{u,k}}
	\begin{split}
	   v_{u,k'} &= \sum_{s_{k'} \in \mathbb{S}} s_{k'}^2 p_{0,k'}\left( s_{k'};\btheta_{u,k'} \right) - \mu_{u,k'}^2\\
	   &=\! \frac{ \left(s^{\left(0\right)}\right)^2 + \sum_{\ell=1}^{L-1}\left(s^{\left(\ell\right)}\right)^2 \exp\braces{d_{k',\ell} + \theta_{u,k',\ell}} }{1+\sum_{\ell=1}^{L-1}\exp\braces{ d_{k',\ell} + \theta_{u,k',\ell} }  } \!-\!  \mu_{u,k'}^2,
	\end{split}
\end{equation}
\end{subequations}
respectively.
Consequently, the mean and covariance of the discrete random vector $\bg_{u,k'}s_{k'}, k' \neq k$, in \eqref{equ:Yn} are 
\begin{subequations}
	\begin{equation}
		\Exp\braces{\bg_{u,k'}s_{k'}} = \bg_{u,k'}\mu_{u,k'},
	\end{equation}
	\begin{equation}
		\Var{\bg_{u,k'}s_{k'}} = v_{u,k'}\bg_{u,k'}\bg_{u,k'}^T,
	\end{equation}
\end{subequations} 
respectively.
Then, we can readily obtain that the mean and covariance  of $\overline{\by}_{u,k},  u\in \setposi{U}, k\in\setposi{2K}$, are
\begin{subequations}
	\begin{equation}\label{equ:Exp of overline y_u,k}
		\Exp\braces{\overline{\by}_{u,k}} = \left( \sum_{k' = 1, k'\neq k}^{2K}\bg_{u,k'}\mu_{u,k'} + \bg_{u,k}s_k\right)  \in \bbR^{N_u},
	\end{equation}
\begin{equation}\label{equ:variance of Y_{n,k}}
	\mathbb{V}\braces{\overline{\by}_{u,k}} = \left( \sum_{k'=1, k'\neq k}^{2K}v_{u,k'}\bg_{u,k'}\bg_{u,k'}^T + \sigma_z^2\bI\right)  \in \bbR^{N_u\times N_u},
\end{equation}
\end{subequations}
respectively.
And we have the following theorem.
\begin{theorem}\label{the:CLT of Y_{n,k}}
	If the condition \eqref{equ:condition on v_{n,k}} holds for $\overline{\by}_{u,k}$ in \eqref{equ:Yn}, 
	then $\overline{\by}_{u,k}$ converges in distribution to a real Gaussian random vector $\tilde{\by}_{u,k}$, as $2K$ goes to infinity, i.e.,
	\begin{equation}
			\overline{\by}_{u,k}\overset{d}{\to}\tilde{\by}_{u,k}  \sim \mathcal{N}\left(\Exp\braces{\overline{\by}_{u,k}},\mathbb{V}\braces{\overline{\by}_{u,k}}\right).
	\end{equation}
\end{theorem}

\begin{figure*}
	\begin{equation}\label{equ:condition on v_{n,k}}
	\lim\limits_{K\to \infty} \left( \sum_{k'\neq k, k' = 1}^{2K}\Exp\braces{ \norm{ \left( \mathbb{V}\braces{\overline{\by}_{u,k}}\right)^{-\frac{1}{2}} \bg_{u,k'}\left( s_{k'} - \mu_{u,k'} \right)  }^3 } + \Exp\braces{ \norm{ \left( \mathbb{V}\braces{\overline{\by}_{u,k}}\right)^{-\frac{1}{2}} \left( \bw_{u,k}' - \bg_{u,k}s_k \right)  }^3  } \right) = 0
\end{equation}
\hrule
\end{figure*}

From Theorem \ref{the:CLT of Y_{n,k}}, we can obtain that when $2K$ is large and the condition \eqref{equ:condition on v_{n,k}} approximately holds, the PDFs of $\overline{\by}_{u,k}$ and $\tilde{\by}_{u,k}$ are approximately equivalent.
In one of the simplest cases, assuming that $\mathbb{V}\braces{\overline{\by}_{u,k}}$ is a diagonal covariance matrix, condition \eqref{equ:condition on v_{n,k}} holds as long as the variance of each component of $\bg_{u,k'}s_{k'}$, $k' \neq k$, in \eqref{equ:Yn} does not tend to zero as $K$ tends to infinity. 
This ensures that $\bg_{u,k'}s_{k'}$, $k' \neq k$, is random rather than deterministic,  which is necessary for the application of the Berry-Esseen theorem.
By replacing $f\left(\overline{\by}_{u,k}\right)$ with $f_{\textrm{G}}\left(\overline{\by}_{u,k};\Exp\braces{\overline{\by}_{u,k}},\mathbb{V}\braces{\overline{\by}_{u,k}}   \right)$ in \eqref{equ:kappa app 1},
we can obtain
\begin{equation}\label{equ:kappa}
   \kappa_{u,k}\left(s_k,\by_{u}\right) \propto f_{\textrm{G}}\left(\overline{\by}_{u,k};\Exp\braces{\overline{\by}_{u,k}},\mathbb{V}\braces{\overline{\by}_{u,k}}   \right)\Big|_{\overline{\by}_{u,k} = \by_u}.
\end{equation}
Combining \eqref{equ:marginals of pn} and \eqref{equ:kappa}, we can obtain \eqref{equ:pn sk app},
\begin{figure*}
	\begin{equation}\label{equ:pn sk app}
			\begin{split}
			&\ p_{u,k}\left(s_k;\btheta_{u}\right)\\
			\propto  &\ \mathrm{exp}\braces{ \sum_{\ell=1}^{L-1} \left(d_{k,\ell} + \theta_{u,k,\ell} \right) \delta\left( s_k - s^{\left(\ell\right)} \right) - \frac{\bg_{u,k}^T\left( \mathbb{V}\braces{\overline{\by}_{u,k}}  \right)^{-1}\bg_{u,k}}{2}\left( s_k - \frac{\bg_{u,k}^T\left( \mathbb{V}\braces{\overline{\by}_{u,k}
					} \right)^{-1}\ba_{u,k}}{\bg_{u,k}^T\left( \mathbb{V}\braces{\overline{\by}_{u,k}}  \right)^{-1}\bg_{u,k}}  \right)^2  }
		\end{split}
	\end{equation}
	\hrule
\end{figure*}
where 
\begin{equation}\label{equ:a_{u,k}}
	\begin{split}
		\ba_{u,k} &\triangleq \by_{u} - \sum_{k' = 1, k' \neq k}^{2K} \bg_{u,k'}\mu_{u,k'} \\
		&= \by_{u} - \bG_u\bmu_{u} + \bg_{u,k}\mu_{u,k} \in \bbR^{N_u},
	\end{split}
\end{equation}
\begin{equation*}
	\bmu_{u} \triangleq \left[\mu_{u,1}, \mu_{u,2}, \cdots,\mu_{u,2K} \right]^T \in \bbR^{2K},
\end{equation*}
$s_k \in \mathbb{S}$, $k\in \setposi{2K}$, $u \in  \setposi{U}$,
and the derivation is given in Appendix \ref{App:A}.
Substituting $s_k = s^{\left(\ell\right)}$ into \eqref{equ:pn sk app}, we can obtain
\begin{subequations}\label{equ:values of marginals of pu}
	\begin{align}
		& \ p_{u,k}\left(s_k;\btheta_{u}\right)\big|_{s_k = s^{\left(0\right)}}
		 \nonumber\\
		=&\ C_{u,k}\exp\braces{  -\frac{\left( s^{\left(0\right)} -  \tilde{\mu}_{u,k}  \right)^2}{2 \tilde{v}_{u,k} }  },\\ 
		&\ p_{u,k}\left(s_k;\btheta_{u}\right)\big|_{s_k = s^{\left(\ell\right)}}  \nonumber\\
		 =&\ C_{u,k}\exp\braces{d_{k,\ell} + \theta_{u,k,\ell}  -\frac{\left( s^{\left(\ell\right)} -  \tilde{\mu}_{u,k}  \right)^2}{2 \tilde{v}_{u,k}    }  },\label{equ:pro of pn ell}
	\end{align}
\end{subequations}
where 
\begin{subequations}\label{equ:inter variables}
	\begin{equation}
		\tilde{v}_{u,k} \triangleq \frac{1}{\bg_{u,k}^T\left( \Var{\overline{\by}_{u,k}}\right)^{-1} \bg_{u,k} },
	\end{equation}
	\begin{equation}
				\tilde{\mu}_{u,k} 
			\triangleq \tilde{v}_{u,k} \bg_{u,k}^T\left( \mathbb{V}\braces{\overline{\by}_{u,k}
			} \right)^{-1}\ba_{u,k},
	\end{equation}
\end{subequations}
$C_{u,k}$ is the normalization factor,
and $\ell \in \setposi{L-1}$ in \eqref{equ:pro of pn ell}.
Consequently, according to the definition in \eqref{equ:def of mp}, the relationship in \eqref{equ:mp solution} and \eqref{equ:values of marginals of pu}, the EACS of the $m$-projection $p_0\left(\bs;\btheta_{0u}\right)$ can be  calculated as
\begin{subequations}\label{equ:mp solution2}
		\begin{equation}
		\btheta_{0u} = \left[ \btheta_{0u,1}^T,\btheta^T_{0u,2},\ldots,\btheta^T_{0u,2K} \right]^T,	
	\end{equation}
	\begin{equation}
		\btheta_{0u,k} = \left[ \theta_{0u,k,1}, \theta_{0u,k,2}, \ldots, \theta_{0u,k,L-1} \right]^T,
	\end{equation}
	\begin{equation}
		\theta_{0u,k,\ell} = \frac{\left( s^{\left(0\right)}-s^{\left(\ell\right)} \right)\left[ \left(s^{\left(0\right)} + s^{\left(\ell\right)}\right)-2\tilde{\mu}_{u,k} \right]}{2\tilde{v}_{u,k}}
		+\theta_{u,k,\ell}, 
	\end{equation}
\end{subequations}
where $u\in \setposi{U}$, $k\in \setposi{2K}$, and $\ell \in \setposi{L-1}$.

We next discuss the efficient implementation of the approximate calculation.
In the approximate calculation, the calculation of the inversions of $\braces{\mathbb{V}\braces{\overline{\by}_{u,k}}|u \in \setposi{U}, k \in \setposi{2K}}$ in \eqref{equ:inter variables} is the most complex.
Direct calculations of these inversions will traverse both the subscripts $u$ and $k$, and thus introduce a total of $2UK$ matrix inversions of $N_u\times N_u$ dimension.
Given $U$ and $K$, we next reduce the total number of inversions to $U$ by the means of Sherman-Morrison formula.
Recalling the definition of $\mathbb{V}\braces{\overline{\by}_{u,k}}$ in \eqref{equ:variance of Y_{n,k}}, we have 
\begin{equation}
	\begin{split}
		\mathbb{V}\braces{\overline{\by}_{u,k}} 
		= \bG_u\Diag{\bv_u}\bG_u^T + \sigma_z^2\bI - v_{u,k}\bg_{u,k}\bg_{u,k}^T,
	\end{split}
\end{equation}
where $\bG_u \in \bbR^{N_u \times 2K}$ is defined in \eqref{equ:G_u}, $\bv_u$ is defined as 
\begin{equation}\label{equ:v_u}
	\bv_u \triangleq \left[v_{u,1}, v_{u,2}, \cdots, v_{u,2K}   \right]^T \in \bbR^{2K},
\end{equation}
and $v_{u,k}$ is given in \eqref{equ:v_{u,k}}. According to Sherman-Morrison formula, we can obtain 
\begin{equation}\label{equ:Sher-Mo formula}
	\begin{split}
		&\left( \mathbb{V}\braces{\overline{\by}_{u,k}} \right)^{-1} \\
		=&\bA_u + \frac{v_{u,k}}{1 - v_{u,k}\bg_{u,k}^T\bA_u\bg_{u,k}}\bA_u\bg_{u,k}\left( \bA_u\bg_{u,k} \right)^T,
	\end{split}
\end{equation}
where
\begin{equation}\label{equ:Au}
	\bA_u \triangleq \left(\bG_u\Diag{\bv_u}\bG_u^T + \sigma_z^2\bI\right)^{-1} \in \bbR^{N_u \times N_u}
\end{equation}
is symmetric.
In \eqref{equ:Sher-Mo formula}, we can find that $\bA_u$ only varies with $u$, and hence we only need to compute $U$ inverse matrices of $N_u\times N_u$ dimension to obtain
\begin{equation*}
	\braces{\left( \Var{\overline{\by}_{u,k}} \right)^{-1}| u \in \setposi{U}, k\in \setposi{2K}}.
\end{equation*}

Now, we discuss the calculation of $\bA_u$.
To obtain the inversion in \eqref{equ:Au}, we can either follow \eqref{equ:Au} directly, or, based on the Woodbury identity, in the  following way:
\begin{equation}\label{equ:Au another}
	\bA_u = \frac{1}{\sigma_z^2}\bI - \frac{1}{\sigma_z^4}\bG_u\left( \textrm{Diag}^{-1}\braces{\bv_u} + \frac{1}{\sigma_z^2}\bG_u^T\bG_u \right)^{-1}\bG_u^T.
\end{equation}
The computational complexities (the number of real-valued multiplications) of \eqref{equ:Au} and \eqref{equ:Au another} are $\mathcal{O}\left(P\right)$ and $\mathcal{O}\left(Q\right)$, respectively, where 
\begin{subequations}\label{equ:P and Q}
	\begin{equation}
		P = N_u^3 + 2KN_u^2,
	\end{equation}
	\begin{equation}
		Q = 8K^3 + 8K^2N_u + 2KN_u^2.
	\end{equation}
\end{subequations}
In practice, we use the one with a lower complexity to get $\bA_u$.
We summarize the approximate calculation of the $m$-projection as Algorithm \ref{Alg:MP}.
It is not difficult to check that the IGA for signal detection in \cite{IGADE} is a special case of GIGA with $U = 2N_r$.
\begin{algorithm}[t]
	\SetAlgoNoLine 
	\caption{Approximate Calculation of the $m$-Projection from $p_u\left(\bs;\btheta_{u}\right)$ onto the OBM}
	\label{Alg:MP}
	\KwIn{Sub-vector $\by_{u}$, sub-matrix $\bG_u$, number of users $K$,  dimension $N_u$ of $\by_{u}$, the alphabet $\mathbb{S} = \braces{s^{\left(0\right)},s^{\left(1\right)}, \ldots,s^{\left( L-1 \right)}}$, the NP $\bd$, the EACS $\btheta_{u}$ of $p_u\left(\bs;\btheta_{u}\right)$, the noise variance $\sigma_z^2$.}	
    1. Calculate $\braces{\mu_{u,k}}_{k=1}^{2K}$ and $\braces{v_{u,k}}_{k=1}^{2K}$ as \eqref{equ:mean and variance of marginals of p0}; \\
    2. Calculate $\braces{\ba_{u,k}}_{k=1}^{2K}$ as \eqref{equ:a_{u,k}};\\
    3. Calculate $\bv_u$ as \eqref{equ:v_u};\\
    4. Calculate $P$ and $Q$ as \eqref{equ:P and Q};\\
    5. \textbf{if} $P \le Q$ \textbf{then}  \\
        \textcolor{white}{5. \textbf{if}} Calculate $\bA_u$ as  \eqref{equ:Au}; \\
    \textcolor{white}{5.} \textbf{else}\\
        \textcolor{white}{5. \textbf{if}} Calculate $\bA_u$ as  \eqref{equ:Au another}; \\
    \textcolor{white}{5.} \textbf{end if}\\   
    6. Calculate $\left( \mathbb{V}\braces{\overline{\by}}_{u,k} \right)^{-1}$, $k \in \setposi{2K}$, as \eqref{equ:Sher-Mo formula}; \\
    7. Calculate $\braces{\tilde{\mu}_{u,k}}_{k=1}^{2K}$ and $\braces{\tilde{v}_{u,k}}_{k=1}^{2K}$ as \eqref{equ:inter variables};\\
    8. Calculate EACS $\btheta_{0u}$ as \eqref{equ:mp solution2};\\
	\KwOut{The EACS $\btheta_{0u}$ of the $m$-projection $p_0\left(\bs;\btheta_{0u}\right)$.}
\end{algorithm}

\subsection{Computational Complexity}

We now give the computational complexity of the two ways of calculating the $m$-projection $p_0\left(\bs;\btheta_{0u}\right)$.
In this work, we use the number of real-valued multiplications as the measure for computational complexity.
The complexity of direct calculation is $\mathcal{O}\left( L^{2K} \right)$ since we can only obtain the marginal probabilities of $p_u\left(\bs;\btheta_{u}\right)$ by means of an exhaustive search.
When the modulation order and the number of users are large, direct calculation will be unaffordable for practical.
According to the steps of Algorithm \ref{Alg:MP}, we can obtain that the computational complexity of approximately calculating the $m$-projection from $p_u\left(\bs;\btheta_{u}\right)$ onto the OBM is 
\begin{equation*}
		\mathcal{O}\left(\min\left( P,Q \right) + 6KN_u^2 + 4KL\right),
\end{equation*}
where $P$ and $Q$ are given by \eqref{equ:P and Q}, $K$ is the number of users, $N_u =\frac{2N_r}{U}$, $N_r$ is the number of antennas at BS, $L = \sqrt{\tilde{L}}$, and $\tilde{L}$ is the modulation order.
Consequently, the computational complexity of GIGA with approximate calculation of $m$-projection is $\mathcal{O}\left(C_U\right)$ per iteration, where 
\begin{equation}\label{equ:Comp}
	C_U =  U\min\left( P,Q \right) + 24K\frac{N_r^2}{U}   + 4KUL,
\end{equation}
$U$ is the number of subsets. 
As $U$ increases from $1$, $C_U$ decreases. However, when $U$ is greater than a certain threshold, $C_U$ will start to increase. We will see this observation in the simulation results below. 
This also shows that the computational complexity of GIGA with $U = 2N_r$ (which is equivalent to IGA in \cite{IGADE})  is not necessarily the lowest compared to the cases when $U$ takes other values.

\section{Simulation Results}
This section provides simulation results to illustrate the performance of GIGA in ultra-massive MIMO system. 
The uncoded BER is used to measure the detection performance in the simulations. 
For generating the channel matrix, we employ the widely-used QuaDRiGa \cite{quadriga}.
In QuaDRiGa, the BS consists of a uniform planar array (UPA) with $N_r = N_{r,v}\times N_{r,h}$ antennas, where $N_{r,v}$ and $N_{r,h}$ are the numbers of the antennas at each vertical column and horizontal row, respectively.
The BS is positioned at coordinates $\left( 0,0,25 \right)$.
Users are randomly distributed within a $120^{\circ}$ sector with a radius of $r = 200$m around $(0, 0, 1.5)$.
Our results are averaged for $1000$ channel matrix realizations.
We summarize the main simulation parameters in Table \ref{tab:para}. 
\begin{table}[htbp]
	\centering
	\caption{Parameter Settings of the Simulation}\label{tab:para}
	\begin{tabular}{cc}
		\hline
		Parameter &Value \\
		\hline
		Number of BS antennas $N_{r,v}\times N_{r,h}$ & $16\times 64$ \\
		UT number $K$ & $240$ \\
		Center frequency $f_c$ & $4.8$GHz \\
		Simulation scenario &  3GPP\_38.901\_UMa\\
		Modulation Mode & QAM \\
		Modulation Order $\tilde{L}$ & $4$, $16$, $64$ \\
		Number of subsets $U$ & $16$, $128$, $512$, $2048$\\
		\hline
	\end{tabular}
\end{table}
Each user's average transmitted symbol power is normalized to $1$.
The \textrm{SNR} is defined as $\textrm{SNR} = {K}/{\tilde{\sigma}_z^2}$.
Based on the received signal model \eqref{equ:rece model},
we compare the proposed GIGA with the following detectors.\\
\textbf{LMMSE}: The classic LMMSE detector with hard-decision, 
\begin{equation}
	\bs_{\textrm{MMSE}} = \left( \bG^T\bG + \sigma_z^2\bI \right)^{-1}\bG^T\by,
\end{equation}
where the hard-decision is performed as
\begin{equation}
	s_{k,\textrm{MMSE}} = \argmin{s_k \in \mathbb{S}}\left| s_k - \left[ \bs_{\textrm{MMSE}} \right]_k \right|^2, k\in \setposi{2K}.
\end{equation} 
The computational complexity of the LMMSE detector is $\mathcal{O}\left(8\left(2N_rK^2 + K^3 \right) \right)$ \cite{6841617}.
\\
\textbf{EP}: The EP detector proposed in \cite{6841617}, where the hard-decision is also performed.
Its computational complexity is $\mathcal{O}\left( 8\left(N_rK^2 + K^3\right) \right)$ per iteration \cite{6841617}.\\
\textbf{AMP}: The classic Bayesian inference algorithm AMP \cite{AMP}.
AMP could approximately calculate the \textsl{a posteriori} marginals.
Then, the MPM detection (\eqref{equ:MPM dete}) is performed.
The computational complexity of AMP is $\mathcal{O}\left( 8\left(N_rK \right) \right)$ per iteration \cite{AMP}.

\subsection{BER Performance}

\begin{figure}[htbp]
	\centering
	\includegraphics[width=0.5\textwidth]{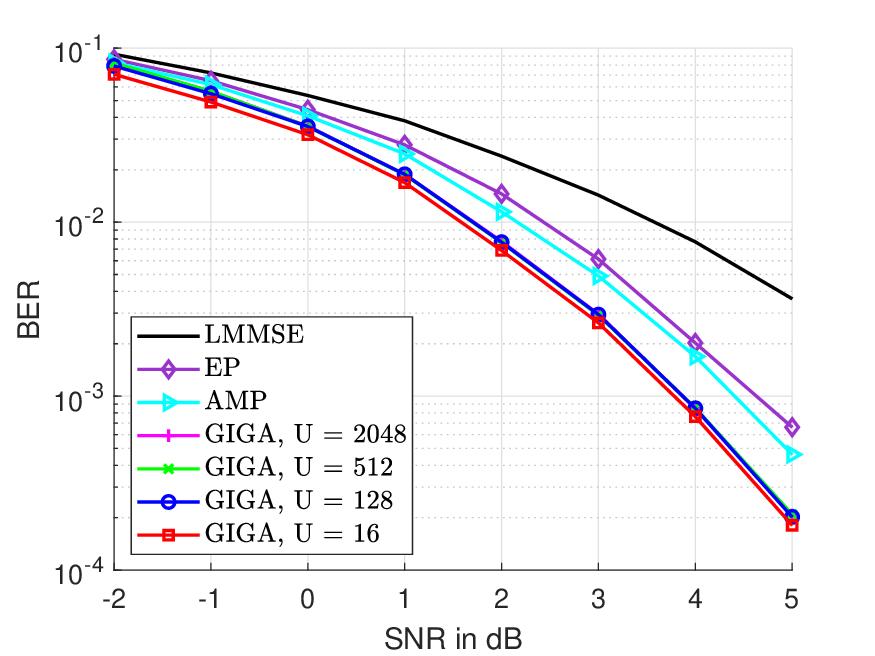}
	\caption{ BER performance of GIGA compared with AMP, EP and LMMSE under $4$-QAM.}
	\label{fig:SNR_4QAM}
\end{figure}

\begin{figure}[htbp]
	\centering
		\includegraphics[width=0.5\textwidth]{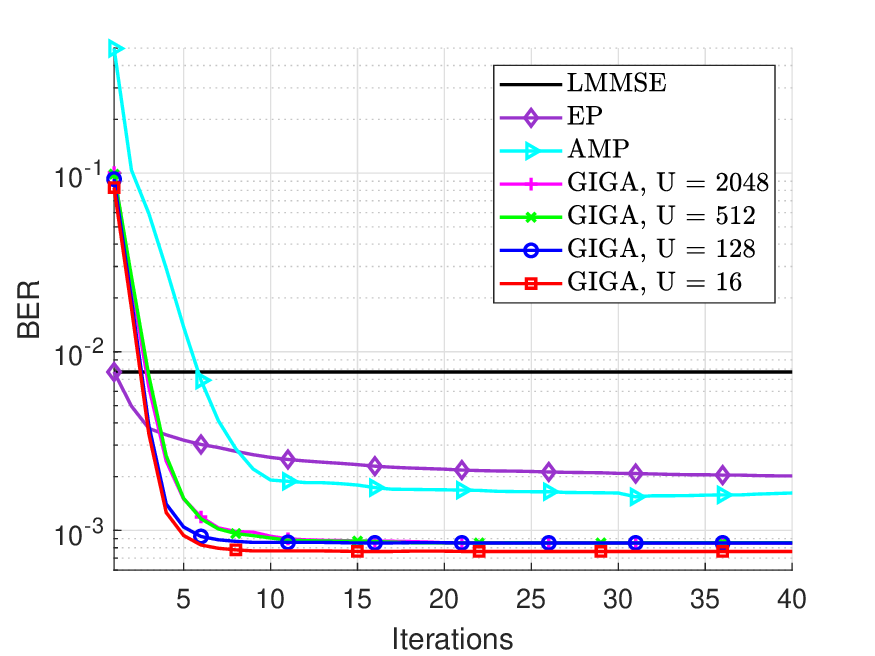}
			\caption{ Convergence performance of GIGA compared with EP and AMP at SNR = $4$ dB under $4$-QAM.}
		\label{fig:It_4QAM}
\end{figure}

We begin with $4$-QAM modulation.
The BER performances of different algorithms are shown in Fig. \ref{fig:SNR_4QAM}.
The iteration numbers of GIGA with $U = 16$, $U = 128$, $U = 512$ and $U = 2048$ are set to be $7$, $10$, $15$ and $15$, respectively.
The iteration numbers of AMP and EP are set to be $30$ and $40$, respectively.
In Fig. \ref{fig:SNR_4QAM}, the BER performance of all the iterative algorithms outperform that of LMMSE detector.
For GIGA, the difference in BER performance between different numbers of subsets is small.
At a BER of $10^{-3}$, the SNR gains of GIGA over AMP and EP are approximately $0.5$dB and $0.7$dB, respectively.
Fig. \ref{fig:It_4QAM} shows the convergence performance of all the iterative algorithms where the SNR is set as $4$dB.
In this case,
with the increase of $U$, the number of iterations for GIGA to converge increases, while the BER performance decreases. 
On the other hand, overall, the performance gap and the difference in the number of iterations  between different $U$ is limited.
AMP and EP converge in around $30$ and $40$ iterations, respectively.
Additionally, it's worth noting that the BER performance of EP with a single iteration is equivalent to that of LMMSE, as the EP detector with one iteration is tantamount to the LMMSE detector \cite{6841617}.

\begin{figure}[htbp]
		\centering
			\includegraphics[width=0.5\textwidth]{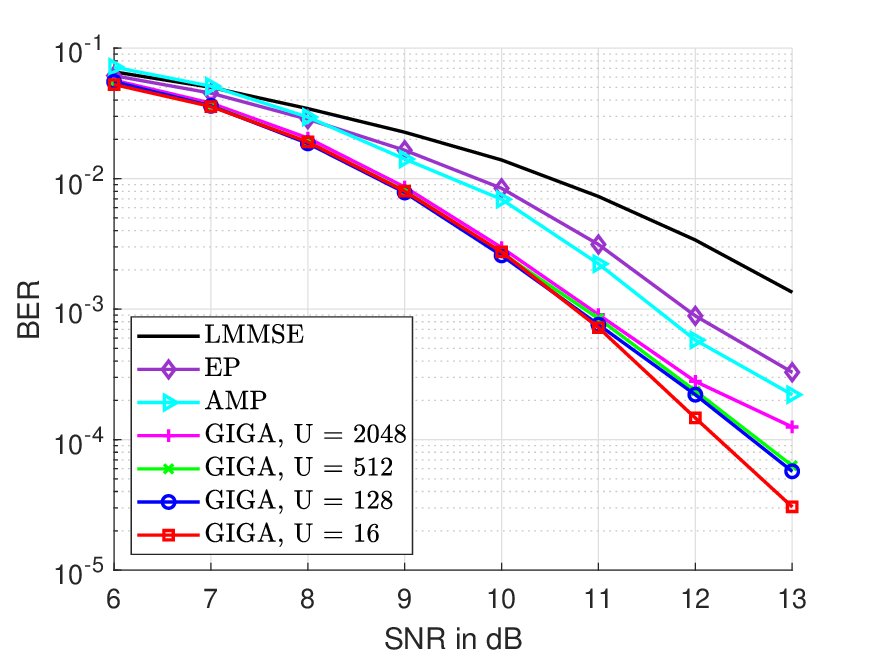}
			\caption{ BER performance of GIGA compared with AMP, EP and LMMSE under $16$-QAM.}
			\label{fig:SNR_16QAM}
\end{figure}

\begin{figure}[htbp]
	\centering
	\includegraphics[width=0.5\textwidth]{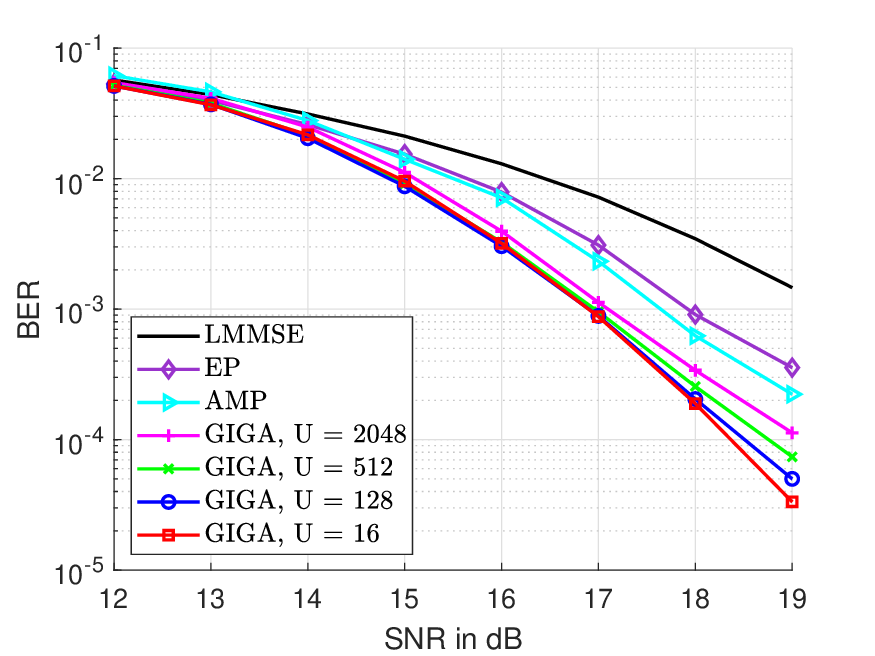}
	\caption{ BER performance of GIGA compared with AMP, EP and LMMSE under $64$-QAM.}
	\label{fig:SNR_64QAM}
\end{figure}

Fig. \ref{fig:SNR_16QAM} and \ref{fig:SNR_64QAM} show the BER performances for $16$-QAM and $64$-QAM, respectively.
In Fig. \ref{fig:SNR_16QAM}, the iteration numbers for GIGA with $U = 16$, $U = 128$, $U = 512$, and $U = 2048$ are set as $15$, $20$, $25$, and $30$, respectively.
The iteration numbers of AMP and EP are set as $50$ and $40$, respectively.
It can be found that GIGA still achieves the best BER performance.
The performance gap between GIGA with different $U$ becomes obvious when SNR is high.
The BER performance of GIGA with $U=16$ at SNR = $12$dB is close to that of GIGA with $U=2048$ at $13$dB.
At a BER of $10^{-3}$, the SNR gains of GIGA over AMP and EP are about $0.7$dB and $0.9$dB, respectively.
For $64$-QAM, the iteration numbers of GIGA with $U = 16$, $U = 128$, $U = 512$, and $U = 2048$ are set as $15$, $25$, $30$, and $35$, respectively.
The iteration numbers of AMP and EP are set as $70$ and $40$, respectively.
Similar observations to those in Fig. \ref{fig:SNR_16QAM} can be obtained.
The BER performance of GIGA with $U=16$ at SNR = $18$dB is close to that of GIGA with $U=2048$ at $18.5$dB.
At a BER of $10^{-3}$, the SNR gains of GIGA over AMP and EP are around $0.7$dB and $1$dB, respectively.

\begin{figure}[htbp]
	\centering
	\includegraphics[width=0.5\textwidth]{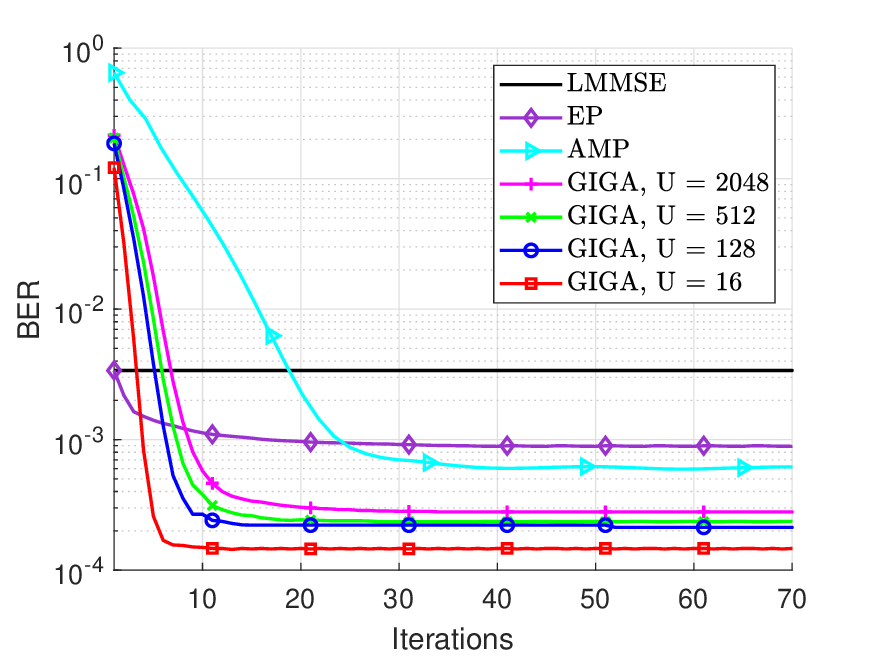}
	\caption{ Convergence performance of GIGA compared with EP and AMP at SNR = $12$ dB under $16$-QAM.}
	\label{fig:It_16QAM}
\end{figure}

\begin{figure}[htbp]
	\centering
	\includegraphics[width=0.5\textwidth]{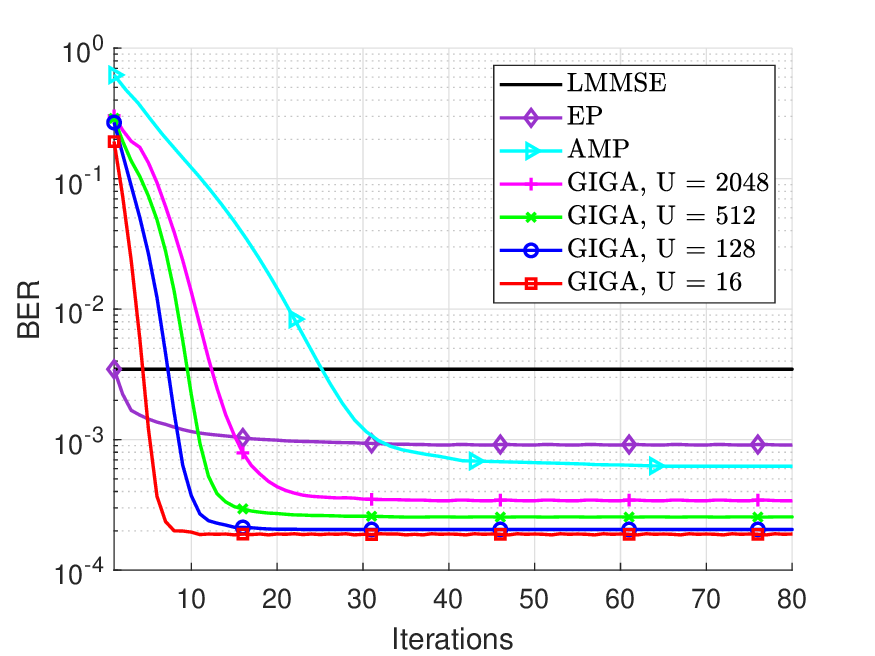}
	\caption{ Convergence performance of GIGA compared with EP and AMP at SNR = $18$ dB under $64$-QAM.}
	\label{fig:It_64QAM}
\end{figure}

Fig. \ref{fig:It_16QAM} and \ref{fig:It_64QAM} show the convergence performances of all the iterative algorithms under $16$-QAM and $64$-QAM, respectively.
From Fig. \ref{fig:It_16QAM}, we can find that in the scenario with $16$-QAM and SNR = $12$dB, the smaller the number of subsets in GIGA, the higher its convergence rate and the lower BER that can be obtained.
The disparity between different $U$ is much more pronounced than that with $4$-QAM.
GIGA with $U = 16$, $U = 128$, $U = 512$, and $U = 2048$ require about $10$, $15$, $20$, and $30$ iterations to converge, respectively.
AMP and EP converge in around $50$ amd $35$ iterations, respectively.
From Fig. \ref{fig:It_64QAM}, we observe that in the scenario with $64$-QAM and SNR of $18$dB,  disparity between GIGA with different $U$ is still observable.
GIGA with $U = 16$ converges within $15$ iterations while GIGA with $U = 128$, $512$, and $2048$ converge in around $25$, $35$, and $50$ iterations, respectively.
AMP and EP require about $60$ and $30$ iterations to converge,  respectively.

\subsection{Complexities}
\begin{figure}[htbp]
	\centering
	\includegraphics[width=0.5\textwidth]{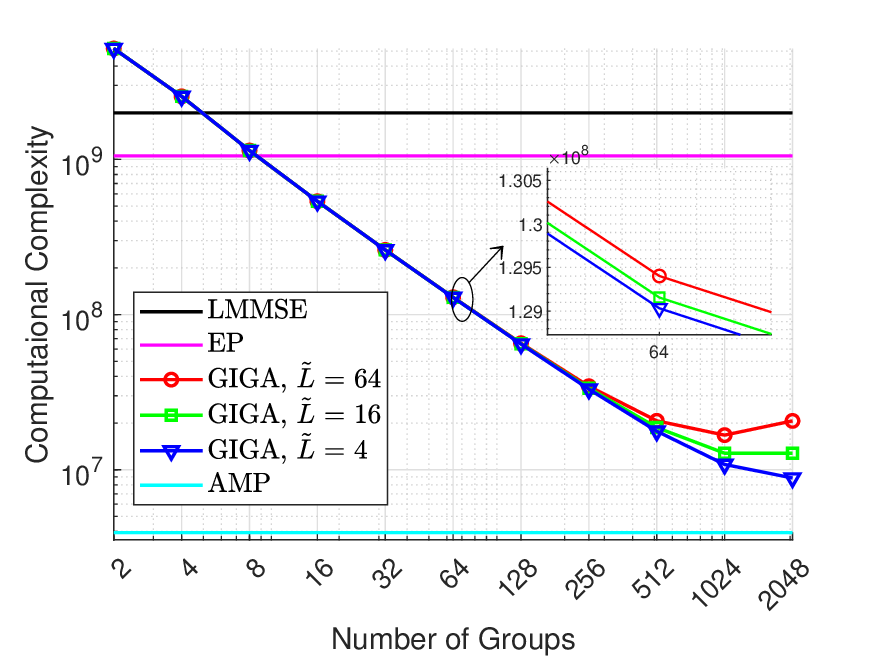}
	\caption{ Complexities of different algorithms versus the number of subsets.}
	\label{fig:complexity}
\end{figure}

The computational complexities of different algorithms are plotted in Fig. \ref{fig:complexity}.
The x-axis is the number of subsets in GIGA.
The numbers of iterations of GIGA, EP and AMP are all set to be $1$.
Among all the iterative algorithms, AMP has the lowest computational complexity.
The complexity of GIGA decreases gradually as the number $U$ of subsets  increases.
Also, we can find a special case when the modulation order is $64$, its computational complexity at $U = 2048$ is sightly higher than that of $U = 1024$.
When $U > 8$, the computational complexity of GIGA becomes lower than that of EP. 
The gap between the two increases rapidly with the number of subsets.

We now discuss the overall computational complexities of different algorithms in our simulations.
Among all algorithms, the complexity of EP is the highest.
Although its complexity at each iteration is lower than that of the LMMSE detection, we can see from Figs. \ref{fig:It_4QAM}, \ref{fig:It_16QAM} and \ref{fig:It_64QAM} that EP requires about tens of iterations to converge, which leads to its highest overall computation.
In our simulations, the number of subsets for GIGA is set to be $16$, $128$, $512$ and $2048$, respectively.
From Figs. \ref{fig:It_4QAM}, \ref{fig:It_16QAM} and \ref{fig:It_64QAM}, we can find that GIGA with $U = 16$ converges within ten iterations, while under other $U$, it converges in tens of iterations.
Under this condition, the overall computational complexity of GIGA with $U = 16$ is comparable to that of the LMMSE detection, while under other $U$, its overall computational complexity is lower than that of the LMMSE detection.
Still, AMP has the lowest overall computational complexity.

\begin{figure}[htbp]
	\centering
	\includegraphics[width=0.5\textwidth]{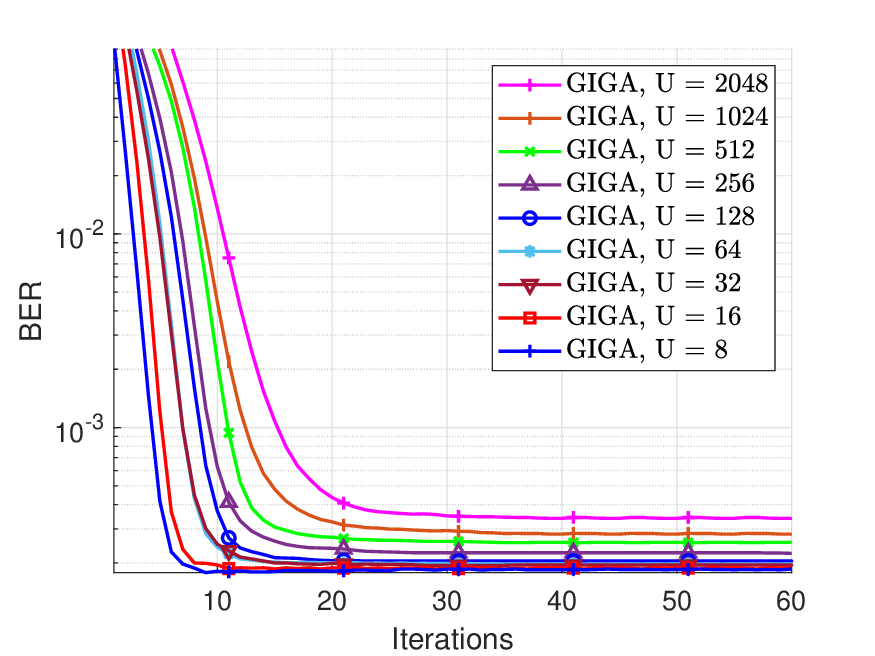}
	\caption{ Convergence performance of GIGA with multiple $U$ at SNR = $18$ dB under $64$-QAM.}
	\label{fig:It_64QAM_MU}
\end{figure}

\begin{figure}[htbp]
	\centering
	\includegraphics[width=0.5\textwidth]{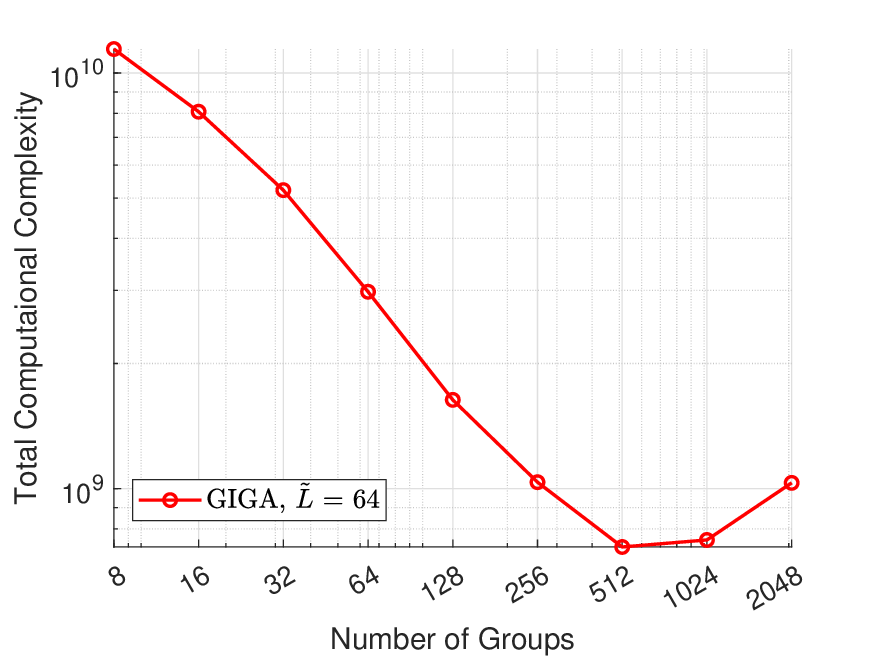}
	\caption{ Total Complexities of GIGA versus the number of subsets under $64$-QAM.}
	\label{fig:complexityMU}
\end{figure}

To further illustrate the overall computational complexity of GIGA with different number $U$ of subsets, we first plot the convergence performance of GIGA with more $U$ in Fig. \ref{fig:It_64QAM_MU}, where the modulation is $64$-QAM and SNR $=18$dB.
Then, Fig. \ref{fig:complexityMU} illustrates the overall computational complexity of GIGA with different $U$.
For each $U$, the overall computational complexity is defined as $T_{U}C_U$, where $T_{U}$ is the number of iterations required for GIGA  to converge when the number of subsets is $U$, and $C_U$ is the complexity of its single iteration.
From Fig. \ref{fig:complexityMU}, we can find that increasing the number of subsets does not necessarily reduce the overall computational complexity. 
Larger $U$ brings smaller single iteration complexity, but it also brings more iterations.
In the case with $64$-QAM and SNR $=18$dB, the overall computational complexity is lowest when $U = 512$, and the gap between its BER performance and the best BER performance brought by $U = 8$ is relatively small.

\section{Conclusion}
In this paper, we have proposed GIGA for ultra-massive MIMO systems. 
We frame the signal detection as an MPM detection problem. 
Leveraging information geometry theory, our objective is to compute approximations of the \textsl{a posteriori} marginals of the transmitted symbols.
Through grouping the components of the received signal $\by$, we factorize the \textsl{a posteriori} probability distribution.
On this basis, we define the AMs, where each AM is related to one group of components of $\by$.
Then, we calculate the approximations of the \textsl{a posteriori} marginals through the $m$-projections from the distributions of AMs onto the OBM.
We give a direct calculation of the $m$-projection as well as an approximate calculation based on the Berry-Esseen theorem.
Efficient implementation of approximate calculation is also discussed. 
Simulation results demonstrate that GIGA achieves the best BER performance within a limited number of iterations compared to existing approaches. 
This showcases the potential of GIGA as an efficient and potent detector in ultra-massive MIMO systems.
As a final remark, although this paper only considers the case when all the groups have the same size, it is straightforward to generalize to the case when the groups have different sizes.


\begin{figure*}[t]
	\begin{equation}\label{equ:aux in App A}
		\begin{split}
			\Theta_{u,k} &= \left( \bg_{u,k}s_k - \ba_{u,k} \right)^T\left( \mathbb{V}\braces{\overline{\by}_{u,k}}  \right)^{-1}\left( \bg_{u,k}s_k - \ba_{u,k} \right) 
			\overset{\left(\textrm{a}\right)}{\propto}  \bg_{u,k}^T\left( \mathbb{V}\braces{\overline{\by}_{u,k}}  \right)^{-1}\bg_{u,k}s_k^2 - 2\bg_{u,k}^T\left( \mathbb{V}\braces{\overline{\by}_{u,k}} \right) ^{-1}\ba_{u,k}s_k \\
			&\overset{\left(\textrm{b}\right)}{\propto}  \bg_{u,k}^T\left( \mathbb{V}\braces{\overline{\by}_{u,k}}  \right)^{-1}\bg_{u,k}\left( s_k - \frac{\bg_{u,k}^T\left( \mathbb{V}\braces{\overline{\by}_{u,k}
				} \right)^{-1}\ba_{u,k}}{\bg_{u,k}^T\left( \mathbb{V}\braces{\overline{\by}_{u,k}}  \right)^{-1}\bg_{u,k}}  \right)^2
		\end{split}
	\end{equation}
	\hrule
\end{figure*}

\appendices

\section{Calculation of \eqref{equ:pn sk app}}\label{App:A}
From \eqref{equ:marginals of pn} and \eqref{equ:kappa}, we can obtain  
	\begin{equation}\label{equ:aux0 in App A}
		p_{u,k}\left(s_k;\btheta_{u}\right) \overset{\left(\textrm{a}\right)}{\propto} \exp\braces{\left(\bd_k + \btheta_{u,k} \right)^T\bt_k - \frac{1}{2} \Theta_{u,k} },
	\end{equation}
where $\left(\textrm{a}\right)$ is obtained by adding the constant independent with $s_k$ and $\by_{u}$, and
	\begin{equation}\label{equ:Theta_u,k}
	\Theta_{u,k} = \left( \by_{u} - \Exp\braces{\overline{\by}_{u,k}} \right)^T\left( \mathbb{V}\braces{\overline{\by}_{u,k}}  \right)^{-1} \left( \by_{u} - \Exp\braces{\overline{\by}_{u,k}} \right).	
\end{equation}
Combining \eqref{equ:Exp of overline y_u,k}, $\left( \by_{u} - \Exp\braces{\overline{\by}_{u,k}} \right)$ in \eqref{equ:Theta_u,k} can be expressed as 
\begin{equation}
	\by_{u} - \Exp\braces{\overline{\by}_{u,k}} \!=\! -\big[ \bg_{u,k}s_k - ( \by_{u} - \sum_{k' = 1, k' \neq k}^{2K} \bg_{u,k'}\mu_{u,k'} )  \big].
\end{equation}
Denote $\ba_{u,k}$ as
\begin{equation*}
 \ba_{u,k} \triangleq	\by_{u} - \sum_{k' = 1, k' \neq k}^{2K} \bg_{u,k'}\mu_{u,k'}.
\end{equation*}
Substituting $\ba_{u,k}$ into \eqref{equ:Theta_u,k},
we can obtain \eqref{equ:aux in App A},
where $\left( \textrm{a}\right)$ comes from removing a constant independent with $s_k$ and $\mathbb{V}\braces{\overline{\by}_{u,k}}$ is symmetric and $\left( \textrm{b}\right)$ comes from adding a constant independent with $s_k$.
Substituting \eqref{equ:aux in App A} into \eqref{equ:aux0 in App A}, we can obtain \eqref{equ:pn sk app}.

\bibliographystyle{IEEEtran}  
\bibliography{IEEEabrv,reference}

\end{document}